\documentclass[11pt,aps,prd,preprint,groupedaddress,tightenlines,nofootinbib,floatfix]{revtex4}
\usepackage[pdftex]{graphicx}

\usepackage{amssymb}
\usepackage{amsmath}
\usepackage{epstopdf}

\makeatletter
\newcommand{\figcaption}[1]{\def\@captype{figure}\caption{#1}}
\newcommand{\tblcaption}[1]{\def\@captype{table}\caption{#1}}
\makeatother

\def\simge{\mathrel{%
       \rlap{\raise 0.511ex \hbox{$>$}}{\lower 0.511ex \hbox{$\sim$}}}}
\def\simle{\mathrel{
       \rlap{\raise 0.511ex \hbox{$<$}}{\lower 0.511ex \hbox{$\sim$}}}}

\begin{document}

\preprint{UTHEP-742, J-PARC-TH-0209}

\title{End point of the first-order phase transition of QCD 
in the heavy quark region by reweighting from quenched QCD
}

\author{Shinji Ejiri$^{1}$, Shota Itagaki$^{2}$, Ryo Iwami$^{3}$, Kazuyuki Kanaya$^{4,5}$, \\
Masakiyo Kitazawa$^{6,7}$, Atsushi Kiyohara$^{6}$, 
Mizuki Shirogane$^{2}$, Takashi Umeda$^{8}$ \\
(WHOT-QCD Collaboration)
}
\affiliation{
$^1$Department of Physics, Niigata University, Niigata 950-2181, Japan\\
$^2$Graduate School of Science and Technology, Niigata University, Niigata 950-2181, Japan\\
$^3$Track Maintenance of Shinkansen, Rail Maintenance 1st Department, East Japan Railway Company Niigata Branch, Niigata, Niigata 950-0086, Japan\\
$^4$Tomonaga Center for the History of the Universe, University of Tsukuba, Tsukuba 305-8571, Japan\\
$^5$Faculty of Pure and Applied Sciences, University of Tsukuba, Tsukuba, Ibaraki 305-8571, Japan\\
$^6$Department of Physics, Osaka University, Toyonaka, Osaka 560-0043, Japan\\
$^7$J-PARC Branch, KEK Theory Center, Institute of Particle and Nuclear Studies, KEK, 203-1, Shirakata, Tokai, Ibaraki 319-1106, Japan\\
$^8$Graduate School of Education, Hiroshima University, Higashihiroshima, Hiroshima 739-8524, Japan
}

\date{February 18, 2020)}

\begin{abstract}
We study the end point of the first-order deconfinement phase transition in two and 2+1 flavor QCD in the heavy quark region of the quark mass parameter space. 
We determine the location of critical point at which the first-order deconfinement phase transition changes to crossover, and calculate the pseudo-scalar meson mass at the critical point.
Performing quenched QCD simulations on lattices with the temporal extents $N_t=6$ and 8, 
the effects of heavy quarks are determined using the reweighting method.
We adopt  the hopping parameter expansion to evaluate the quark determinants in the reweighting factor. 
We estimate the truncation error of the hopping parameter expansion by comparing the results of leading and next-to-leading order calculations, and study the lattice spacing dependence as well as the spatial volume dependence of the result for the critical point.
The overlap problem of the reweighting method is also examined.
Our results for $N_t=4$ and 6 suggest that the critical quark mass decreases as the lattice spacing decreases and increases as the spatial volume increases.
\end{abstract}

\maketitle

\section{Introduction}
\label{sec:intro}

Quantum chromodynamics (QCD) is known to have a rich phase structure as a function of temperature $T$, quark chemical potential $\mu$, and quark masses $m_q$.
The confined phase at low $T$ and small $\mu$ turns into the deconfined phase when $T$ exceeds the transition temperature $T_c$. The nature of this deconfinement phase transition varies as a function of $\mu$ and $m_q$.
From lattice QCD simulations, the transition is an analytic crossover at small $\mu$ around the physical quark masses \cite{Aoki2006we,RBCBi09} but is expected to change to a first-order transition when we increase $\mu$ or vary $m_q$.
The determination of the critical point where the crossover changes to a first-order transition is important in understanding the nature of quark matter in heavy-ion collisions
\cite{crtpt02,crtpt03,crtpt04,dFP03,dFP07,dFP08,Ejiri12,Jin15,Cher19}.

In 2+1 flavor QCD, which contains dynamical up, down and strange quarks, 
there exist two first-order phase transition regions in the quark mass parameter space at $\mu=0$.
One region is located around the light quark limit (chiral limit):
when all three quarks are massless, the spontaneously broken chiral symmetry at low $T$ is recovered by a first-order chiral transition at which confinement is also resolved \cite{PW}.
Many studies have been done to determine the critical mass where the first-order transition terminates in the quark mass parameter space \cite{
dFP03,dFP07,Iwasaki96,JLQCD99,Schmidt01,Christ03,Bernard05,Cheng07,Bazavov17,Jin14}.
The critical mass was found to be close to the physical point and thus a quantitative determination of it is phenomenologically important. 
However, so far, the continuum limit of the critical mass has not been conclusively determined.
Another first-order transition region is located around the heavy quark limit \cite{Ukawa83,DeTar83,Alexandrou99,Saito1,Saito2,Fromm12,Philipsen17}:
when all the quarks are infinitely heavy, QCD is just the pure gauge SU(3) Yang-Mills theory (quenched QCD), which is known to have a weakly first-order deconfinement transition \cite{FOU}.
This first-order transition changes to crossover when the quark mass becomes smaller than a critical value.
Simulations on coarse lattices suggest that the critical quark mass is large.
However, its continuum limit is also not well understood.

In this paper, we study the end point of the first-order deconfinement transition in the heavy quark region of two and 2+1 flavor QCD,
and evaluate the critical quark mass at the end point.
In our previous papers \cite{Saito1,Saito2} we studied the issue 
at $\mu =0$ and finite $\mu$ on a $24^3 \times 4$ lattice. 
From quenched QCD simulations combined with the hopping parameter expansion, 
we found that the first-order deconfinement transition in the heavy quark limit becomes weaker and eventually disappears as the quark mass decreases. 
We determined the location of the critical surface separating the first-order and crossover regions around the heavy quark limit.

We extend these studies using finer lattices with temporal lattice sizes $N_t=6$ and 8, and compare the results with those at $N_t=4$ to discuss the lattice spacing dependence of the critical point.
We also compute the pseudo-scalar meson mass at the critical point.
Although the approximation by the hopping parameter expansion becomes worse as the lattice spacing decreases, i.e., $N_t$ increases, our approach makes it possible, in particular, to investigate phase structures at finite density with low computational costs, as shown in Ref.~\cite{Saito2}.
Hence, it is worth continuing this study by increasing $N_t$ and finding the limitation of this approach.

This paper is organized as follows. 
In the next section, we define our model and introduce the histogram method \cite{Ejiri07,Ejiri13} to find the end point of the first-order phase transition.
Our simulation parameters are given in Sec.~\ref{sec:simu}.
In Sec.~\ref{sec:kappac}, we determine the critical point in two-flavor QCD by adopting a leading order (LO) approximation of the hopping parameter expansion.
The overlap problem in the reweighting method is also discussed.
In Sec.~\ref{sec:kappacNLO}, influence of the next-to-leading order (NLO) terms of the hopping parameter expansion is evaluated to examine the applicability of the hopping parameter expansion in the study of the critical point.
The pseudo-scalar meson mass is computed at the end point of the first-order phase transition in Sec.~\ref{sec:mass}. 
The lattice spacing dependence of the critical quark mass is discussed for two flavor QCD.
We also determine the boundary of the first-order region in the mass parameter plane of $2+1$ flavor QCD in Sec.~\ref{sec:2+1flavor}.
We summarize our results in Sec.~\ref{sec:conclusion}.
In Appendix~\ref{sec:multipoint}, we introduce the multi-point histogram method used in this study.

\section{Formulation}
\label{sec:method}


The action we study consists of the gauge action and the quark action,
\begin{eqnarray}
S = S_g + S_q,
\label{eq:Stot}
\end{eqnarray} 
where the gauge action is the standard plaquette action given by  
\begin{eqnarray}
S_g = -6 N_{\rm site} \,\beta \, \hat{P},
\label{eq:Sg}
\end{eqnarray} 
where $\beta = 6/g^2$ the gauge coupling parameter,
$N_{\rm site} = N_s^3 \times N_t$ the space-time lattice volume,
and $\hat{P}$ the plaquette operator: 
\begin{eqnarray}
\hat{P}= \frac{1}{18 N_{\rm site}} \displaystyle \sum_{x,\,\mu < \nu} 
 {\rm Re \ Tr} \left[ U_{x,\mu} U_{x+\hat{\mu},\nu}
U^{\dagger}_{x+\hat{\nu},\mu} U^{\dagger}_{x,\nu} \right] .
\label{eq:SgP}
\end{eqnarray} 
Here, $U_{x,\mu}$ is the link variable in the $\mu$ direction at site $x$ and 
$x+ \hat\mu$ the next site in the $\mu$ direction from $x$.

For quarks, we adopt the standard Wilson quark action given by 
\begin{eqnarray}
S_q = \sum_{f=1}^{N_{\rm f}} \sum_{x,\,y} \bar{\psi}_x^{(f)} \, M_{xy} (K_f) \, \psi_y^{(f)} ,
\label{eq:Sq}
\end{eqnarray} 
where $M_{xy}$ is the Wilson quark kernel
\begin{eqnarray}
M_{xy} (K_f) &=& \delta_{xy} 
-K_f \sum_{\mu=1}^4 \left[ (1-\gamma_{\mu})\,U_{x,\mu}\,\delta_{y,x+\hat{\mu}} + (1+\gamma_{\mu})\,U_{y,\mu}^{\dagger}\,\delta_{y,x-\hat{\mu}} \right]  
\label{eq:Mxy}
\end{eqnarray} 
where $K_f$ is the hopping parameter for the $f$th flavor. 

\subsection{Histogram and effective potential}

The order parameter of the deconfinement transition of QCD in the heavy quark limit is given by the Polyakov loop defined as
\begin{equation}
\hat\Omega = \frac{1}{3N_s^3}
\displaystyle \sum_{\mathbf{x}} {\rm tr} \left[ 
U_{\mathbf{x},4} U_{\mathbf{x}+\hat{4},4} U_{\mathbf{x}+2 \cdot \hat{4},4} 
\cdots U_{\mathbf{x}+(N_t -1) \cdot \hat{4},4} \right] , 
\label{eq:ploop}
\end{equation}
where $\sum_\mathbf{x}$ is a summation over one time slice. 
The Polyakov loop provides us with an order parameter for the Z(3) center symmetry, which spontaneously breaks down when $T$ exceeds a critical value. 
In our previous studies \cite{Saito1,Saito2,Ejiri13}, 
we found that the histogram of the Polyakov loop is useful in determining the phase structure around the heavy quark limit.\footnote{A similar approach was proposed in Refs.~\cite{Langfeld16,Langfeld12}.}

We focus on the absolute value of the Polyakov loop $|\Omega|$. 
Denoting $K = (K_1, \cdots, K_{N_{\rm f}})$, we define the histogram of $|\Omega|$ by 
\begin{eqnarray}
w( |\Omega|; \beta, K) 
&=& \int {\cal D} U {\cal D} \psi {\cal D} \bar{\psi} \ \delta(|\Omega| - |\hat{\Omega}|) \ e^{- S_g - S_q} \nonumber \\
&=& \int {\cal D} U \ \delta(|\Omega| - |\hat{\Omega}|) \ 
e^{-S_g}\ \prod_{f=1}^{N_{\rm f}} \det M(K_f) .
\label{eq:hist}
\end{eqnarray}
The probability distribution function of $|\Omega|$ is given by 
$Z^{-1}(\beta, K) \, w(|\Omega|;\beta,K)$ , 
where $Z$ is the partition function defined by
\begin{equation}
Z(\beta,K)  = \int\! w(|\Omega|;\beta,K) \, d|\Omega|, 
\end{equation}
and the naive histogram, obtained by just counting the number of configurations with $|\Omega|$ in a simulation, is given by  $N \, Z^{-1}(\beta, K) \, w(|\Omega|;\beta,K)$, where $N$ is the total number of configurations.

When the transition is of first-order, we expect multiple peaks in the histogram. 
We can thus detect the end point of a first-order transition through a change in the shape of the histogram.
For convenience, we define the effective potential of $|\Omega|$ as
\begin{equation}
V_{\rm eff}(|\Omega|; \beta, K) = -\ln w(|\Omega|; \beta, K) + {\rm (const.)},
\end{equation}
where the constant term is fixed later. 
The effective potential has two minima when the histogram has two peaks.

Using the reweighting method, the parameters $\beta$ and $K$ can be changed 
from the simulation point $(\beta_0, K_0)$ with $K_0 = (K_{0;1}, \cdots, K_{0;N_{\rm f}})$.
The probability distribution of $|\Omega|$ at $(\beta, K)$ is given by
\begin{eqnarray}
Z^{-1}(\beta, K)\, w(|\Omega|; \beta, K)
&=& \frac{ \int {\cal D} U \, \delta(|\Omega| - |\hat{\Omega}|) \,
\prod_f \det M(K_f) \, e^{6 \beta N_{\rm site} \hat{P}} }{
 \int {\cal D} U \,\prod_f \det M(K_f) \, e^{6 \beta N_{\rm site} \hat{P}} }
 \nonumber \\
&=& 
\frac{ \left\langle \delta(|\Omega| - |\hat{\Omega}|) \prod_f \frac{ \det M(K_f)}{\det M(K_{0;f})} \, e^{6 (\beta -\beta_0) N_{\rm site} \hat{P}} \right\rangle_{(\beta_0, K_0)} 
}{ \left\langle \prod_f \frac{ \det M(K_f)}{\det M(K_{0;f})} \, e^{6 (\beta -\beta_0) N_{\rm site} \hat{P}} \right\rangle_{(\beta_0, K_0)} } ,
\label{eq:R}
\end{eqnarray}
where 
\begin{equation}
\langle {\cal O}  \rangle_{(\beta_0, K_0)}
=
\frac{ \int {\cal D} U \, {\cal O} \,
\prod_f \det M(K_{0;f}) \, e^{6 \beta_0 N_{\rm site} \hat{P}} }{
 \int {\cal D} U \,\prod_f \det M(K_{0;f}) \, e^{6 \beta_0 N_{\rm site} \hat{P}} }
\end{equation}
 is the expectation value at $(\beta_0, K_0)$. 
In principle, Eq.~(\ref{eq:R}) should enable us to predict the shape of $w(|\Omega|)$ at any $(\beta, K)$. 
However, statistically reliable data on $w(|\Omega|)$ is available only around $|\Omega| \approx \langle |\hat\Omega| \rangle_{(\beta_0,K_0)}$. 
When $\langle |\hat\Omega| \rangle_{(\beta,K)}$ at the target point $(\beta,K)$ 
shifts a lot from $\langle |\hat\Omega| \rangle_{(\beta_0,K_0)}$, 
it is not easy to obtain a reliable prediction about the nature of the vacuum at $(\beta,K)$.
This is the overlap problem, which is severe around a first-order transition point on large lattices.
The overlap problem can be avoided, to some extent, by combining information obtained at several different simulation points. 
For the histogram, the multi-point histogram method is developed based on the reweighting method~\cite{Saito1,Saito2}.
The method is introduced in Appendix~\ref{sec:multipoint}.

\subsection{QCD in the heavy quark region}
\label{sec:heavyQregion}


To investigate the quark mass dependence of the effective potential in the heavy quark region, we evaluate the quark determinant by the hopping parameter expansion in the vicinity of the heavy quark limit $K=0$.
For each flavor, we have 
\begin{eqnarray}
\ln \left[ \frac{\det M(K)}{\det M(0)} \right]
\;=\; \sum_{n=1}^{\infty} \frac{{\cal D}_{n}}{n!}  \, K^{n} , 
\label{eq:tayexp}
\end{eqnarray}
with
\begin{eqnarray}
{\cal D}_n &\equiv& \left[ \frac{\partial^n \ln \det M}{\partial K^n} \right]_{K=0}
\;=\; (-1)^{n+1} (n-1)! \; {\rm tr} 
\left[ \left( M^{-1} \, \frac{\partial M}{\partial K} \right)^n \right]_{K=0}
\nonumber\\
&=&(-1)^{n+1} (n-1)! \; {\rm tr} 
\left[ \left( \frac{\partial M}{\partial K} \right)^n \right], 
\label{eq:derkappa}
\end{eqnarray}
where $(\partial M/\partial K)_{xy}$ is the term following $K_f$ on the right hand side of Eq.~(\ref{eq:Mxy}).
Therefore, the non-vanishing contributions to ${\cal D}_{n}$ are given by Wilson loops and Polyakov-loop-type loops. 

In this study, we compute the leading order (LO) and the next-to-leading order (NLO) terms from the Wilson loops and Polyakov-loop-type loops. 
The LO contribution consists of the smallest Wilson loop, plaquette $\hat{P}$ [defined by Eq.~(\ref{eq:SgP})], and the shortest Polyakov-loop-type loop, the Polyakov loop $\Omega$ [defined by Eq.~(\ref{eq:ploop})]. 
Because of the anti-periodic boundary condition and gamma matrices in the hopping terms, 
up to the NLO contributions, Eq.~(\ref{eq:tayexp}) reads 
\begin{eqnarray}
 \ln \left[ \frac{\mathrm{det}M(K)}{\mathrm{det}M(0) }\right] &=&
  288N_{\mathrm{site}}K^4 \hat{P} + 768N_{\mathrm{site}}K^6
  \left( 3 \hat{W}_{\mathrm{rec}}+6 \hat{W}_{\mathrm{chair}}+2 \hat{W}_{\mathrm{crown}} \right) + \cdots
\nonumber \\
  && \hspace{-25mm}
 +12\times 2^{N_t}N_s^3 K^{N_t} \mathrm{Re} \hat\Omega 
 + 36\times 2^{N_t}N_s^3 N_t K^{N_t+2} \left( 2 \! \sum_{n=1}^{N_t/2-1} \! \mathrm{Re} \hat\Omega_n  
 +  \mathrm{Re} \hat\Omega_{N_t/2} \right) + \cdots , \ 
 \label{eq:hpe_nlo}
\end{eqnarray}
where a constant term does not appear because the left hand side vanishes at $K=0$.
We assume that $N_t$ is an even number.

\begin{figure}[tb]
\begin{center}
\includegraphics[width=8cm]{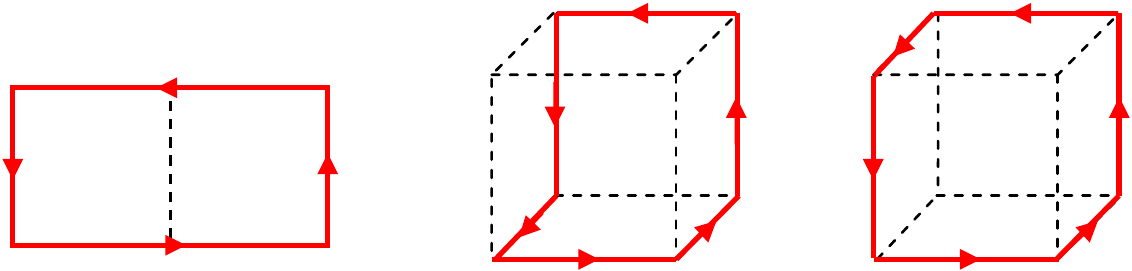}
\end{center}
\caption{Six-step Wilson loops; rectangle (left), chair-type (middle) and crown-type (right).}
\label{fig:6stepwilson}
\end{figure}

\begin{figure}[tb]
\begin{center}
\includegraphics[width=6cm]{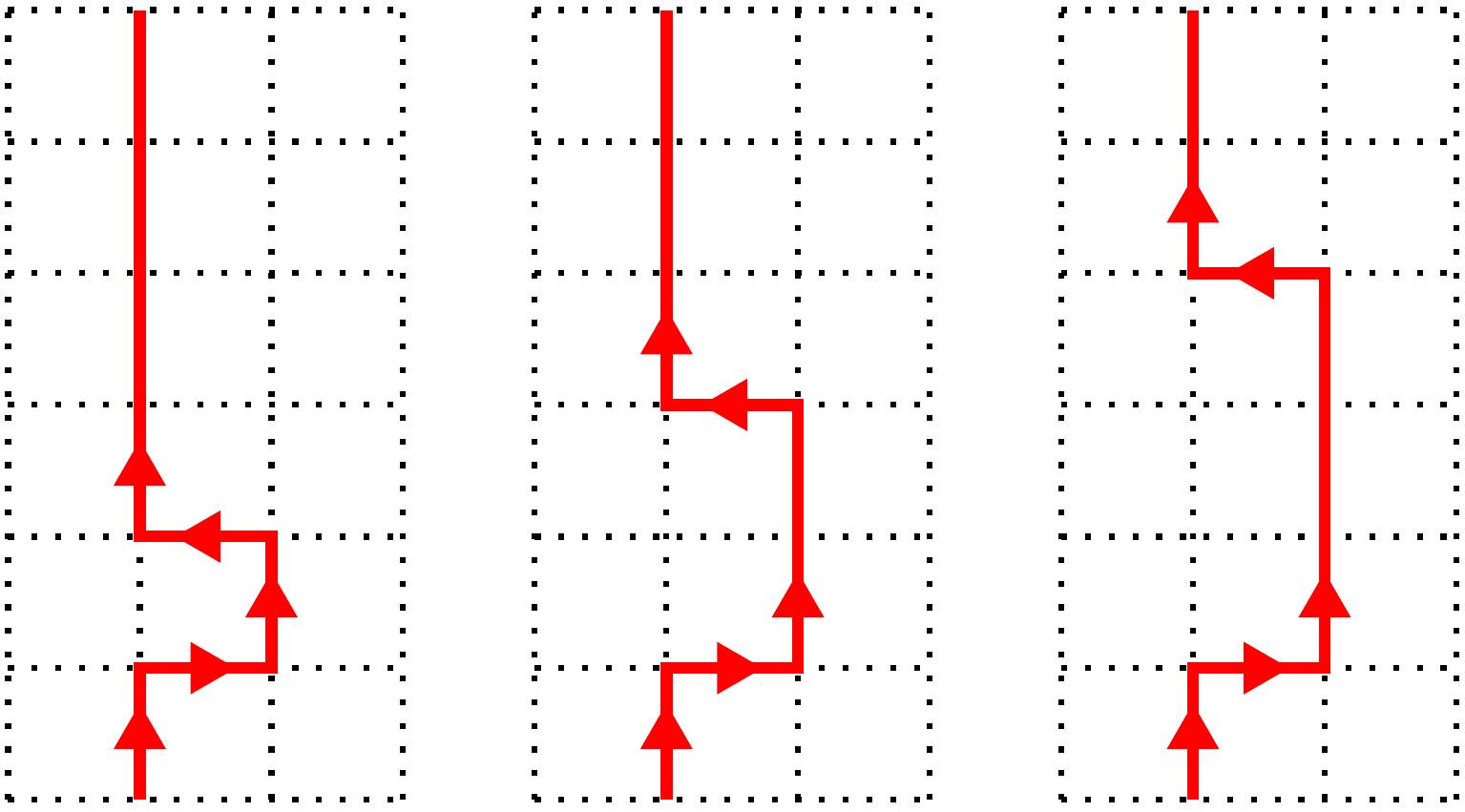}
\end{center}
\caption{$(N_t +2)$-step bent Polyakov loops,  $\Omega_1$ (left), $\Omega_2$ (middle) and $\Omega_3$ (right) for $N_t=6$. The vertical upward direction is the temporal direction.}
\label{fig:bendedpoly}
\end{figure}

The NLO contribution from Wilson loops is given by six-step Wilson loops: rectangle $\hat{W}_{\mathrm{rec}}$, 
chair-type $\hat{W}_{\mathrm{chair}}$, and crown-type $\hat{W}_{\mathrm{crown}}$. 
They are illustrated in Fig.~\ref{fig:6stepwilson}. 
The Polyakov-loop-type loops for the NLO contribution are $(N_t +2)$-step bent Polyakov loops $\hat\Omega_n$ with $n=1, \cdots, N_t/2$, which run one step in a space direction, $n$ steps in the time direction and return to the original line,
e.g., 
\begin{eqnarray}
\hat\Omega_1 &=& \frac{1}{6 N_t N^3_s}\sum_{\mathbf{x}}
\sum_{\mu=\pm1}^{3} 
\frac{1}{3} {\rm tr} \left[ \;
U_{\mathbf{x},\mu} U_{\mathbf{x}+\hat\mu,4} U_{\mathbf{x}+\hat4,\mu}^{\dagger} U_{\mathbf{x}+\hat4,4}
U_{\mathbf{x}+2\cdot\hat4,4} \cdots U_{\mathbf{x}+(N_t-1)\cdot\hat4,4} \right. \nonumber \\
&& 
\left.
+\,U_{\mathbf{x},4} U_{\mathbf{x}+\hat4,\mu} U_{\mathbf{x}+\hat4+\hat\mu,4} U_{\mathbf{x}+2\cdot\hat4,\mu}^{\dagger}
U_{\mathbf{x}+2\cdot\hat4,4} \cdots U_{\mathbf{x}+(N_t-1)\cdot\hat4,4} 
+ \cdots 
\; \right] ,
\label{eq:bendpl1}
\\
\hat\Omega_2 &=& \frac{1}{6 N_t N^3_s} \sum_{\mathbf{x}}
\sum_{\mu=\pm1}^{3} 
\frac{1}{3} {\rm tr} \left[ \;
U_{\mathbf{x},\mu} U_{\mathbf{x}+\hat\mu,4} U_{\mathbf{x}+\hat\mu+\hat4,4}
U_{\mathbf{x}+2\cdot\hat4,\mu}^{\dagger} U_{\mathbf{x}+2\cdot\hat4,4} \cdots U_{\mathbf{x}+(N_t-1)\cdot\hat4,4} 
\right. \nonumber \\ 
&& 
\left.
+\,U_{\mathbf{x},4} U_{\mathbf{x}+\hat4,\mu} U_{\mathbf{x}+\hat\mu+\hat4,4} U_{\mathbf{x}+\hat\mu+2\cdot\hat4,4}
U_{\mathbf{x}+3\cdot\hat4,\mu}^{\dagger} \cdots U_{\mathbf{x}+(N_t-1)\cdot\hat4,4}  
+ \cdots \; \right] .
\label{eq:bendpl2}
\end{eqnarray}
They are illustrated in Fig.~\ref{fig:bendedpoly}.
All of the Wilson loops and Polyakov-loop-type loops are normalized such that 
$\hat{W}_{\mathrm{rec}} = \hat{W}_{\mathrm{chair}} = \hat{W}_{\mathrm{crown}}=1$ and 
$\hat\Omega_n = 1$ in the weak coupling limit, $U_{x,\mu}=1$.

The first term of Eq.~(\ref{eq:hpe_nlo}) is proportional to $\hat{P}$ and thus can be absorbed by a shift $\beta \rightarrow \beta^*$ in the gauge action.
Collecting the contribution from all flavors, we find 
\begin{eqnarray}
\beta^* = \beta + 48 \sum_{f=1}^{N_{\rm f}} K_f^4 .
\label{eq:betastar}
\end{eqnarray}
We note that the six-step Wilson loops are typical operators in improved gauge actions.
Thus, the contributions of these operators can also be absorbed by a shift of improvement parameters of the gauge action.
Because a shift in improvement parameters only affects the amount of lattice discretization errors within the same universality class, the six-step Wilson loop terms will not affect characteristic physical properties of the system in the continuum limit, such as the critical exponents of the phase transition.
In contrast, the terms proportional to the Polyakov-loop-type loops act like external magnetic fields in spin models, 
and thus may change the nature of the phase transition.
Therefore, in our study of NLO contributions, we concentrate on the effects of the Polyakov-loop-type terms on the nature of the phase transition, disregarding the effects of the six-step Wilson loop terms in~Eq.~(\ref{eq:hpe_nlo}).

In Sec.~\ref{sec:kappacLO}, we first study the phase structure only with the LO terms of $O(K^{N_t})$ in $\ln \det M$.
We then study the influence of the NLO terms of $O(K^{N_t+2})$ in~Sec.~\ref{sec:kappacNLO}.
On the other hand, the six-step Wilson loops do affect the detailed properties of the system, such as the values of critical hadron masses.
In Sec.~\ref{sec:mass}, we examine their effects on the pseudo-scalar meson mass by comparing with the results of direct full QCD simulations at $K\ne0$.

\section{Simulation parameters}
\label{sec:simu}

  \begin{table}[tbhp]
   \begin{center}
    \caption{Simulation parameters: lattice size, $\beta$, number of configurations.}
    \begin{tabular}{cccc}
     \hline
     lattice size & $\beta$ & No. of confs. \\ \hline
     $24^3 \times 6$ 
     & 5.8700  & 120000 \\
     & 5.8750  & 176000 \\
     & 5.8800  & 160000 \\
     & 5.8880  & 100190 \\
     & 5.8910  & 80000  \\
     & 5.8930  & 40000 \\ \hline
     $32^3 \times 6$
     & 5.8810  & 100000 \\
     & 5.8850  & 455000 \\
     & 5.8890  & 167000 \\
     & 5.8910  & 149000 \\
     & 5.8950  & 200000 \\
     & 5.9000  & 101000 \\ \hline
     $36^3 \times 6$
     & 5.8910  & 200000 \\
     & 5.8930  & 200000 \\
     & 5.8940  & 200000 \\
     & 5.8949  & 200000 \\ \hline
     $24^3 \times 8$
     & 6.0320  & 18000 \\
     & 6.0380  & 40000 \\
     & 6.0440  & 80000 \\
     & 6.0500  & 80000 \\
     & 6.0560  & 80000 \\
     & 6.0660  & 44700 \\ \hline
    \end{tabular}

    \label{tab:params}
   \end{center}
  \end{table}

  \begin{table}[tbhp]
   \begin{center}
    \caption{Parameter $\Delta$ for the approximate delta function.}
    \begin{tabular}{cc}
     \hline
     lattice size & $\Delta$ \\ \hline
     $24^3 \times 6$ & 0.0026 \\
     $32^3 \times 6$ & 0.003 \\
     $36^3 \times 6$ & 0.003 \\
     $24^3 \times 8$ & 0.0016 \\ \hline
    \end{tabular}
    \label{tab:delta}
   \end{center}
  \end{table}

We calculate the effective potential of $|\Omega|$ in the heavy quark mass region
by performing simulations of quenched QCD on 
$24^3 \times 6$, $32^3 \times 6$, $36^3 \times 6$, and $24^3 \times 8$ lattices.
We generate the quenched configurations using the pseudo heat bath algorithm of SU(3) gauge theory with over relaxation. 
Because the effective potential must be investigated in a wide range of $|\Omega|$ 
to study the nature of the phase transition, 
we perform simulations at several points of $\beta$ and combine these data using the multi-point histogram method. 
Details of the multi-point histogram method are given in Appendix~\ref{sec:multipoint}. 
We mainly investigate on $N_t =6$ lattices. 
The spatial volume dependence is also studied in this case.
 Adjusting the simulation parameters to the critical point, 
the lattice spacing $a$ is given by $a=(N_t T_c)^{-1}$ in terms of $N_t$ and the transition temperature $T_c$.
To study lattice spacing dependence, i.e., $N_t$ dependence, we perform an additional simulation on a lattice with $N_t=8$, and also use the results of $N_t=4$ lattices obtained in Ref.~\cite{Saito1}.
The values of $\beta$ and the number of configurations are summarized in Table~\ref{tab:params}.

To check validity of the reweighting method with the hopping parameter expansion, we also perform full QCD simulations with two favors of dynamical Wilson quarks on $16^3\times32$ lattices, using the hybrid Monte Carlo algorithm.
Simulation parameters for the full QCD simulations are given in Sec.~\ref{sec:mass}.

In numerical evaluation of the histogram defined by Eq.~(\ref{eq:hist}), we adopt a Gaussian approximation for the delta function: 
$\delta(x) \approx \exp [ -(x/\Delta)^2 ] / (\Delta \sqrt{\pi})$.
We choose the width parameter $\Delta$ by considering a balance between the resolution of the histogram and its statistical error.
The values of $\Delta$ we adopt are given in Table~\ref{tab:delta}.
The statistical errors are estimated by the jackknife method.

\section{Critical point of two-flavor QCD in the leading order hopping parameter expansion}
\label{sec:kappac}

\begin{figure}[tb]
\begin{center}
\includegraphics[width=7.5cm]{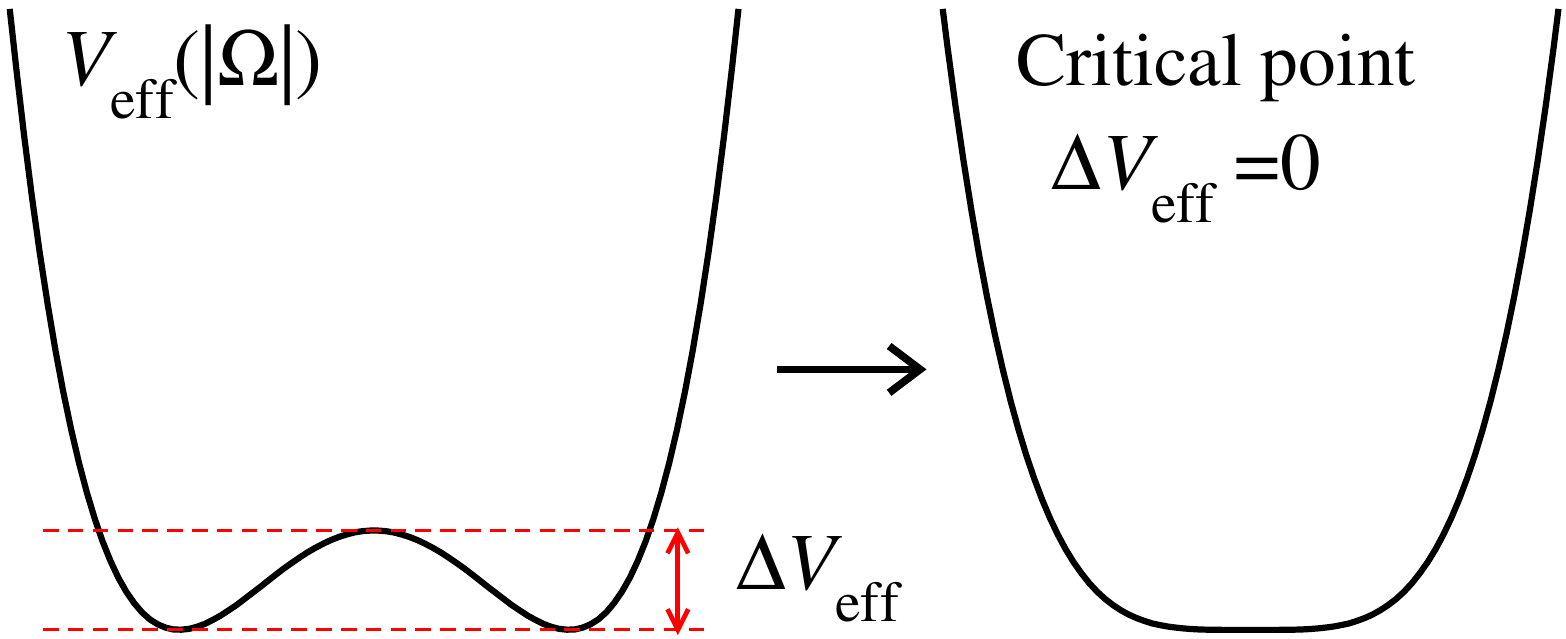}
\end{center}
\caption{Illustration of the effective potential near the critical point, with $\beta$ adjusted to the transition point.}
\label{fig:histogram_veff}
\end{figure}

We investigate how the quark mass affects the shape of the effective potential around the first-order transition point 
using the reweighting method explained in Sec.~\ref{sec:method}.
We first study the case of two-flavor QCD with degenerate $u$ and $d$ quarks.
The extension to the case of $2+1$ flavor QCD is discussed in Sec.~\ref{sec:2+1flavor}.

We calculate the effective potential $V_{\mathrm{eff}}$ at small $K$ by reweighting from $K=0$.
In the first-order transition region, $V_{\mathrm{eff}}$ has two minima.
At each $K$, we first adjust $\beta$ to the transition point by adjusting $\beta$ so that two minimum values of $V_{\mathrm{eff}}$ are equal.
We then measure the difference $\Delta V_{\mathrm{eff}}$ between the peak height in the middle of $V_{\mathrm{eff}}$ and the minimum value as illustrated in Fig.~\ref{fig:histogram_veff}.%
\footnote{In practice, we adopt the average of two minimum values as the minimum value if a slight difference remains between them after the accumulation of full statistics.}
We define the critical point $K_{ct}$, where the first-order transition line terminates, as the value of $K$ where $\Delta V_{\mathrm{eff}}$ vanishes.
The logarithm of the quark determinant $\ln \det M$, Eq.~(\ref{eq:hpe_nlo}), in the reweighting factor is calculated by the LO terms of $O(K^{N_t})$, i.e., Polyakov loop, in this section, 
and the Polykov-loop-type loops up to the NLO terms of $O(K^{N_t+2})$ in Sec.~\ref{sec:kappacNLO}. 
By comparing the results of $K_{ct}$ using only the LO contribution and that using both LO and NLO contributions, we estimate the truncation error of the hopping parameter expansion.

\begin{figure}[tb]
   \begin{minipage}{0.45\hsize}
    \begin{center}
     \includegraphics[width=7.5cm]{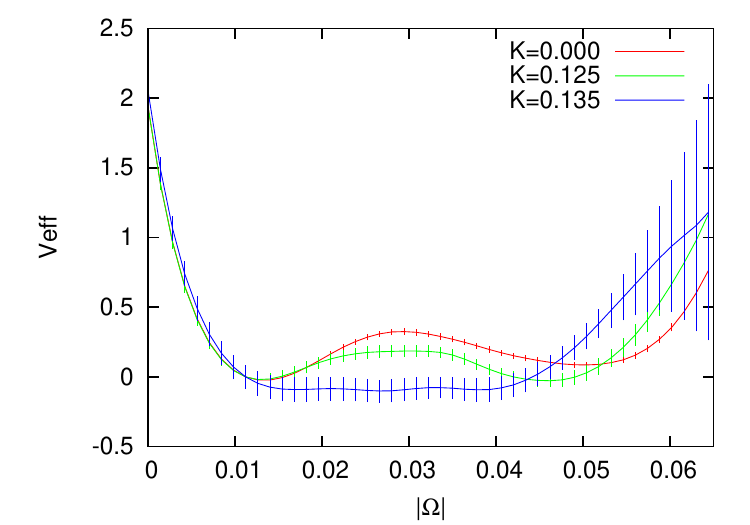}
    \end{center}
    \caption{$V_{\mathrm{eff}}(|\Omega|)$ obtained by the leading order of the hopping parameter expansion on a $24^3\times6$ lattice.}
    \label{fig:24x6veff_lo}
   \end{minipage}
   \hspace{3mm}
   \begin{minipage}{0.45\hsize}
    \begin{center}
     \includegraphics[width=7.5cm]{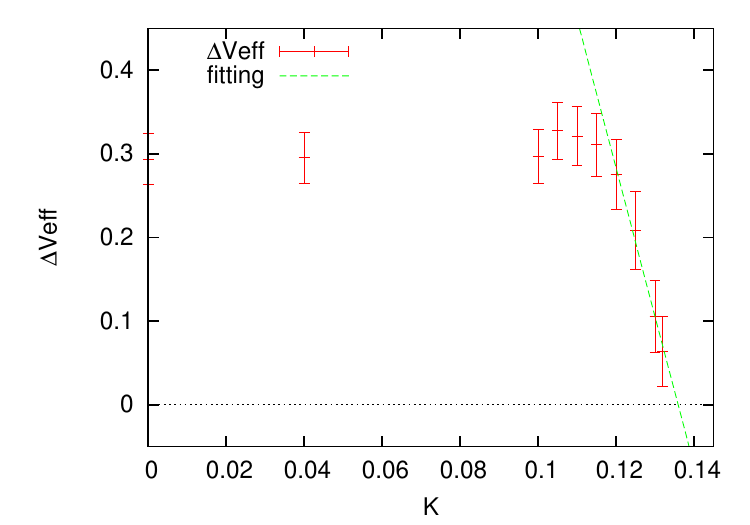}
    \end{center}
    \caption{$\Delta V_{\mathrm{eff}}$ as function of $K$ obtained by the leading order of the hopping parameter expansion 
    on a $24^3\times6$ lattice.}
    \label{fig:24x6dveff_lo}
   \end{minipage}
\end{figure}

 \begin{figure}[tb]
   \begin{minipage}{0.45\hsize}
    \begin{center}
     \includegraphics[width=7.5cm]{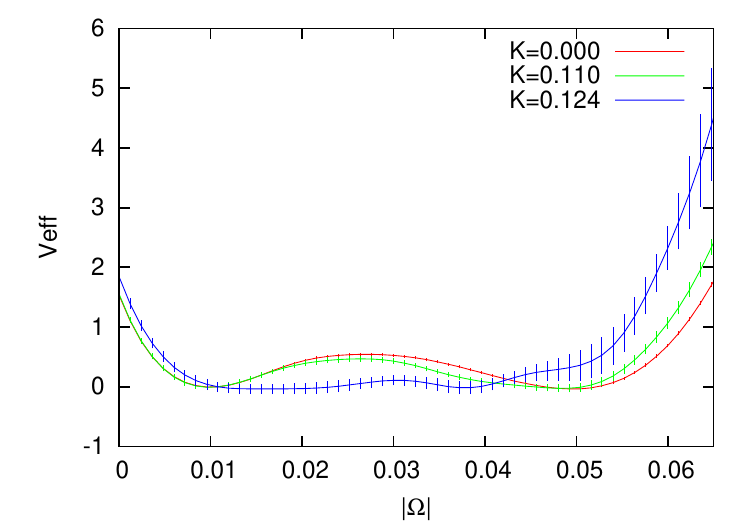}
    \end{center}
    \caption{$V_{\mathrm{eff}}(|\Omega|)$ obtained by the leading order calculation on a $32^3\times6$ lattice.}
    \label{fig:32x6veff_lo}
   \end{minipage}
   \hspace{3mm}
   \begin{minipage}{0.45\hsize}
    \begin{center}
     \includegraphics[width=7.5cm]{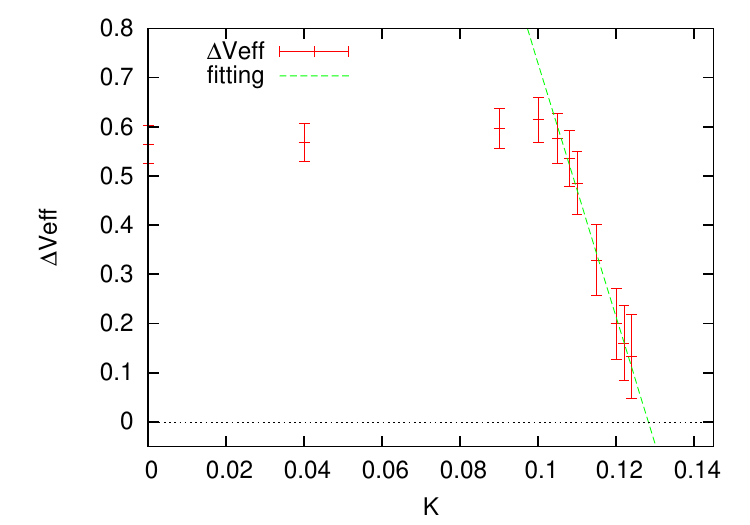}
    \end{center}
    \caption{$\Delta V_{\mathrm{eff}}$ as function of $K$ obtained by the leading order calculation
    on a $32^3\times6$ lattice.}
    \label{fig:32x6dveff_lo}
   \end{minipage}
 \end{figure}

\subsection{Results for $N_t=6$}
\label{sec:kappacLO6}

In Fig.~\ref{fig:24x6veff_lo}, we show the $K$ dependence of $V_{\mathrm{eff}}$ computed on the $24^3 \times 6$ lattice 
using the LO terms of $\ln \det M$.
The effective potential has two minima at $K=0.0$ and 0.125, 
while it becomes almost flat at the minimum at $K=0.135$.
We plot $\Delta V_{\mathrm{eff}}$ as a function of $K$ in Fig.~\ref{fig:24x6dveff_lo}.
Fitting the four smallest data points by a linear function (the dashed line), we obtain the critical point $K_{ct}=0.1359(30)$.

\begin{figure}[tb]
  \begin{center}
   \includegraphics[width=7.5cm]{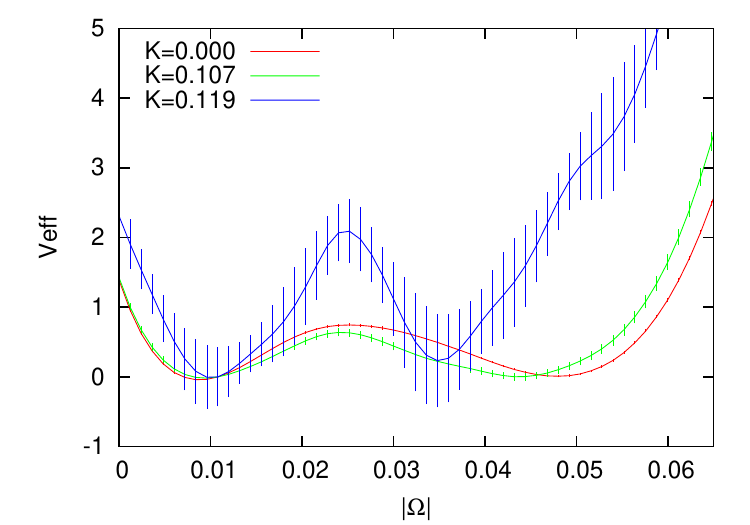}
  \end{center}
  \caption{$V_{\mathrm{eff}}(|\Omega|)$ obtained by the leading order calculation on a $36^3\times6$ lattice.}
  \label{fig:36x6veff_lo}
\end{figure}    

\begin{figure}[tb]
   \begin{minipage}{0.45\hsize}
  \begin{center}
  \includegraphics[width=7.5cm]{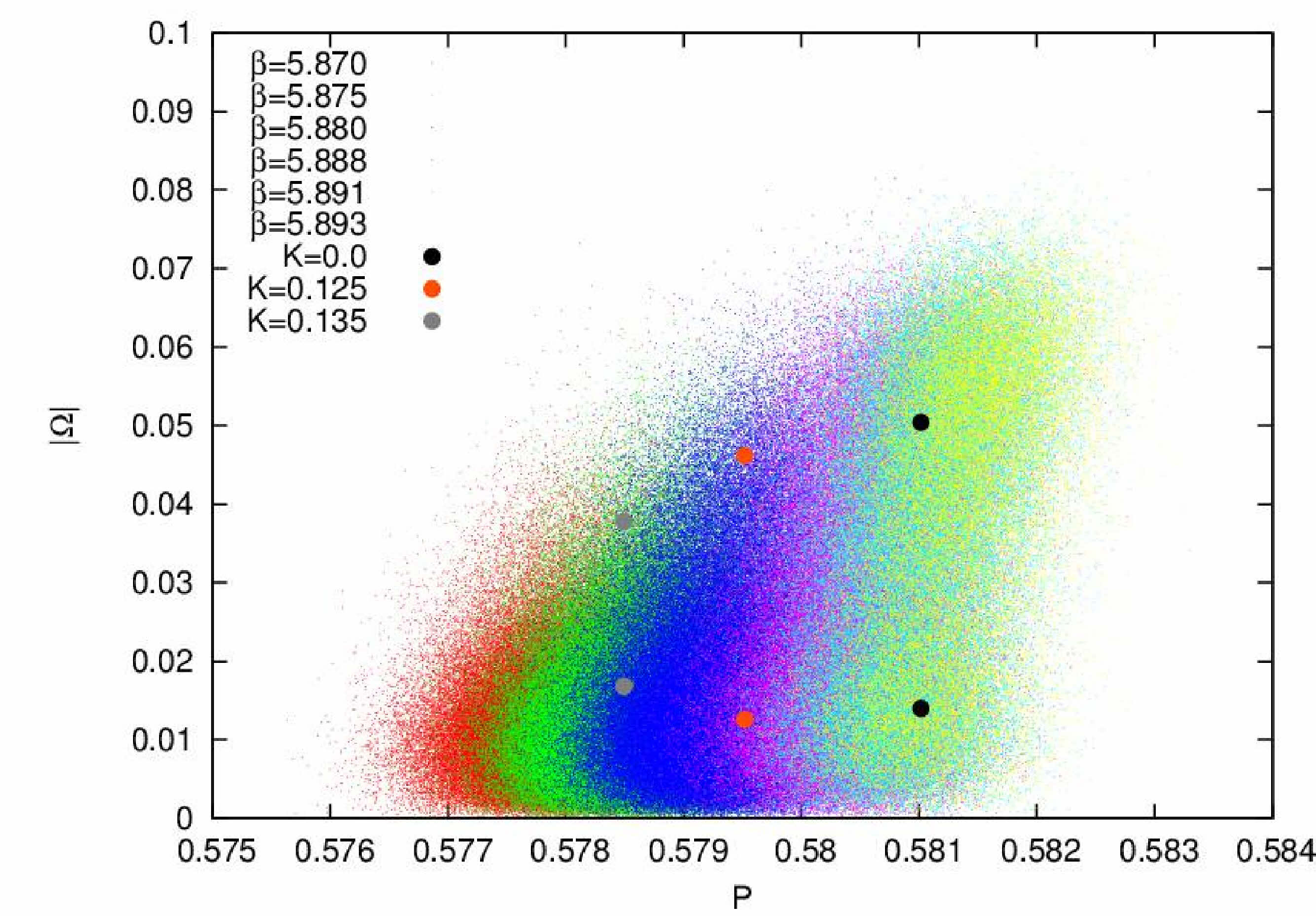}
  \end{center}
  \caption{Distribution of $P$ and $|\Omega|$ on a $24^3\times6$ lattice. 
    The circle symbols are the peak positions on the histogram.}
  \label{fig:24t6_allplot}
   \end{minipage}
   \hspace{3mm}
   \begin{minipage}{0.45\hsize}
  \begin{center}
  \includegraphics[width=7.5cm]{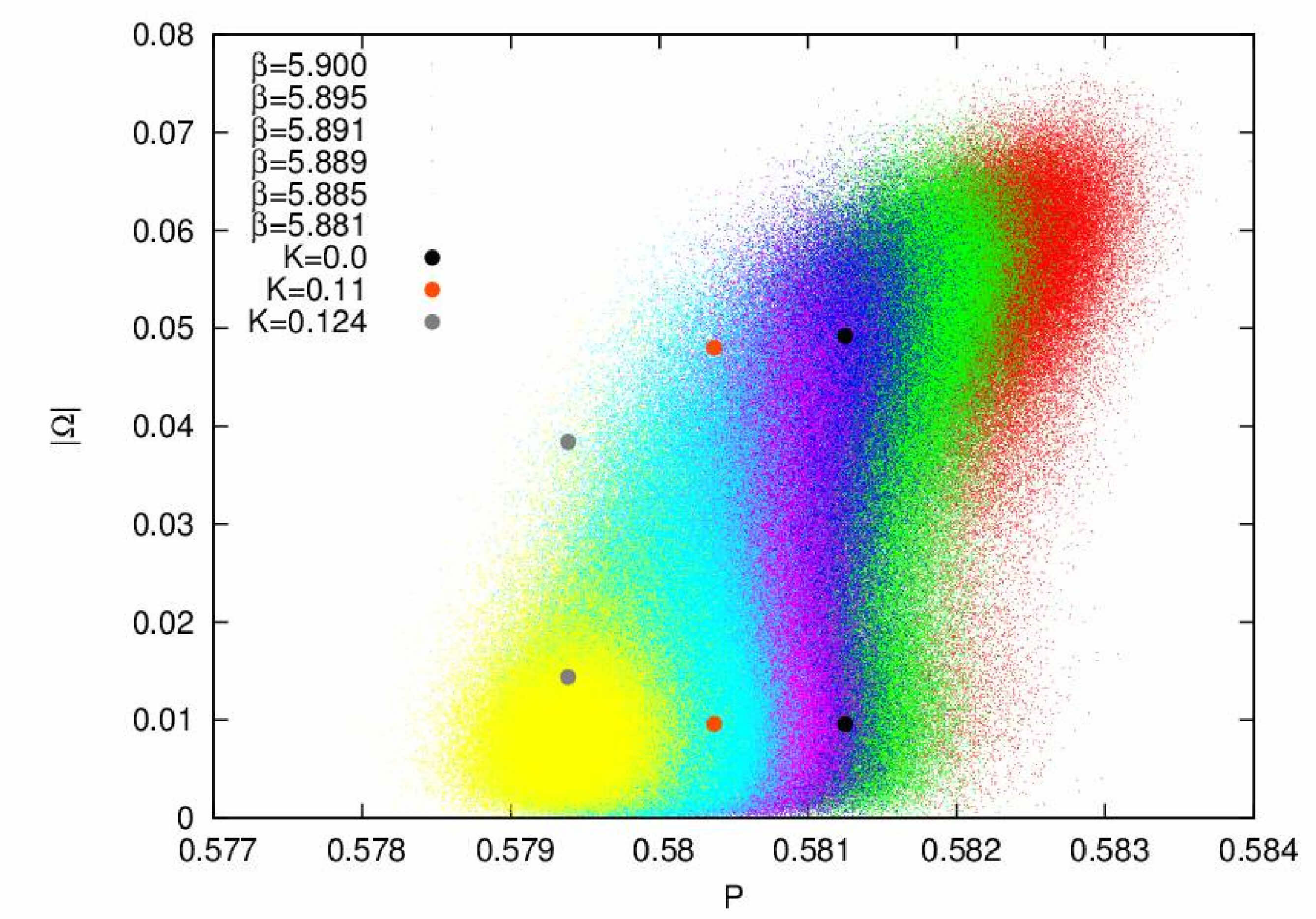}
  \end{center}
  \caption{Distribution of $P$ and $|\Omega|$ on a $32^3\times6$ lattice. 
    The circle symbols are the peak positions on the histogram.}
  \label{fig:32t6_allplot}
  \end{minipage}
\end{figure}

\begin{figure}[tb]
  \begin{center}
  \includegraphics[width=7.5cm]{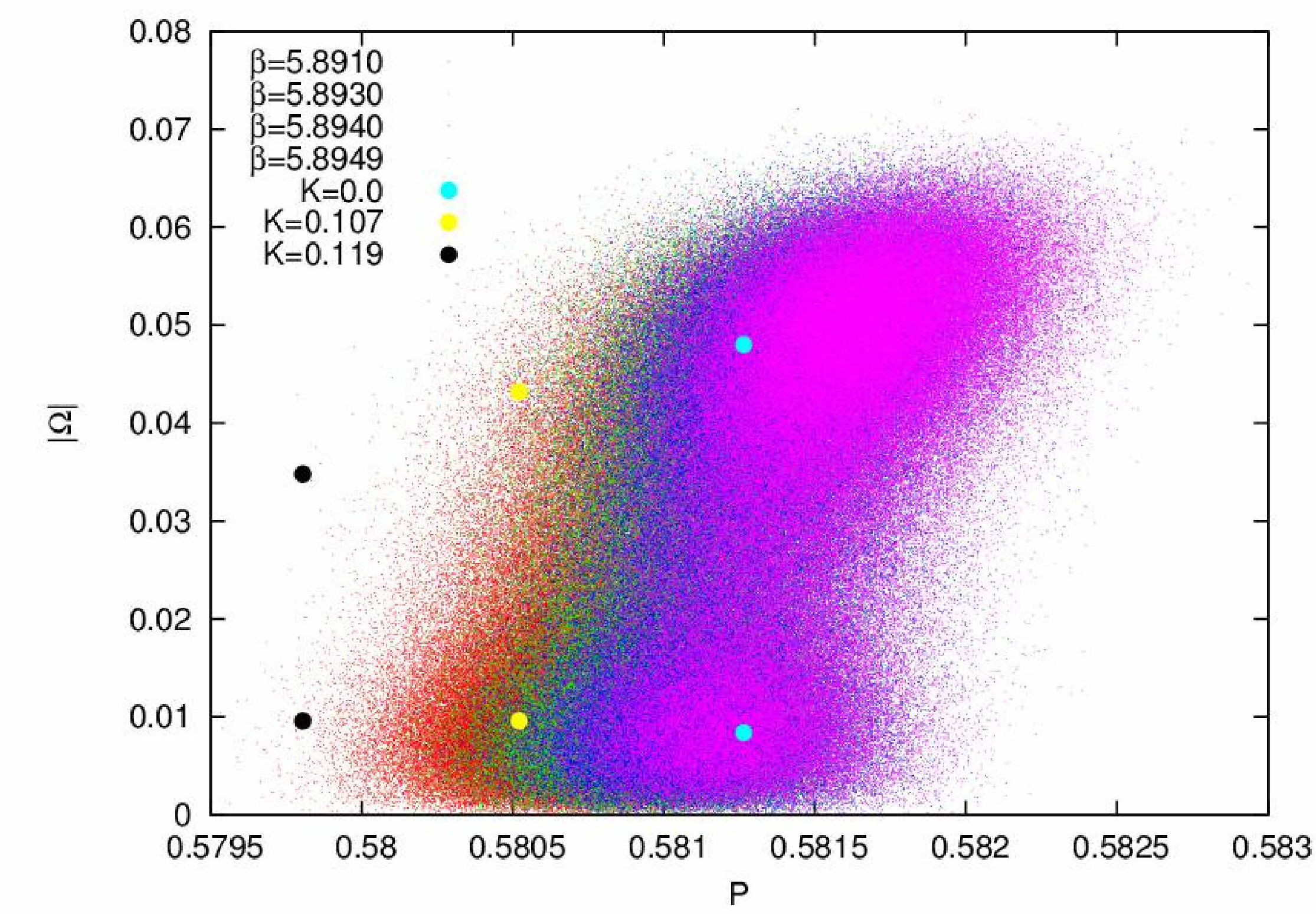}
  \end{center}
  \caption{Distribution of $P$ and $|\Omega|$ on a $36^3\times6$ lattice. }
  \label{fig:36t6_allplot}
\end{figure}    

\begin{figure}[tb]
   \begin{minipage}{0.45\hsize}
  \begin{center}
   \includegraphics[width=7.5cm]{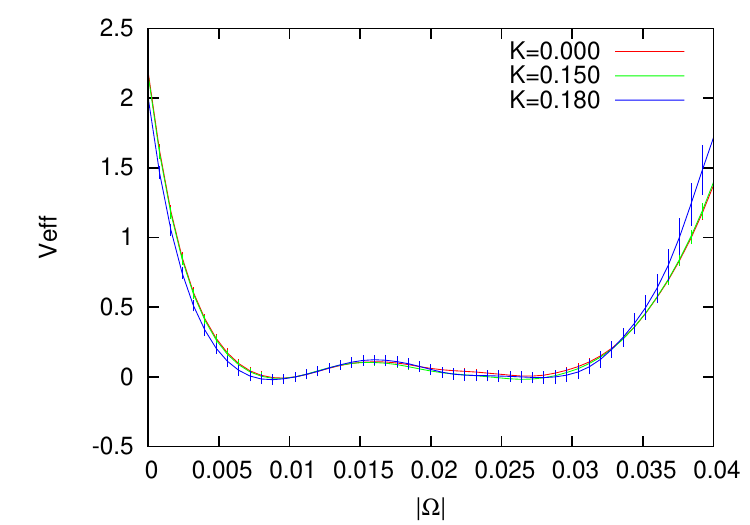}
  \end{center}
  \caption{$V_{\mathrm{eff}}(|\Omega|)$ obtained by the leading order calculation on a $24^3\times8$ lattice.}
  \label{fig:24x8veff_lo}
   \end{minipage}
   \hspace{3mm}
   \begin{minipage}{0.45\hsize}
  \begin{center}
  \includegraphics[width=7.5cm]{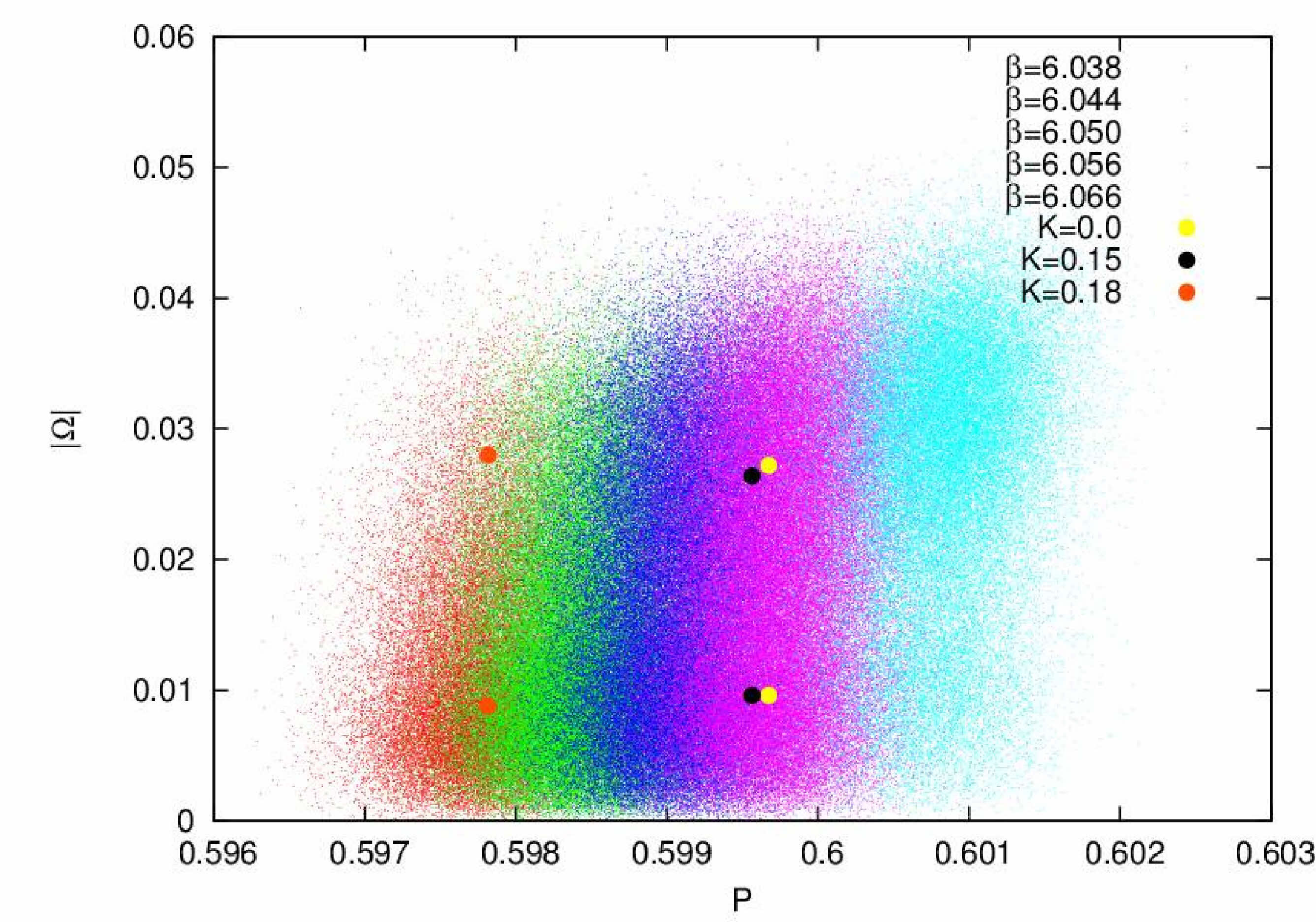}
  \end{center}
  \caption{Distribution of $P$ and $|\Omega|$ on a $24^3\times8$ lattice. }
  \label{fig:24t8_allplot}
  \end{minipage}
\end{figure}

\begin{figure}[tb]
   \begin{center}
   \includegraphics[width=7.5cm]{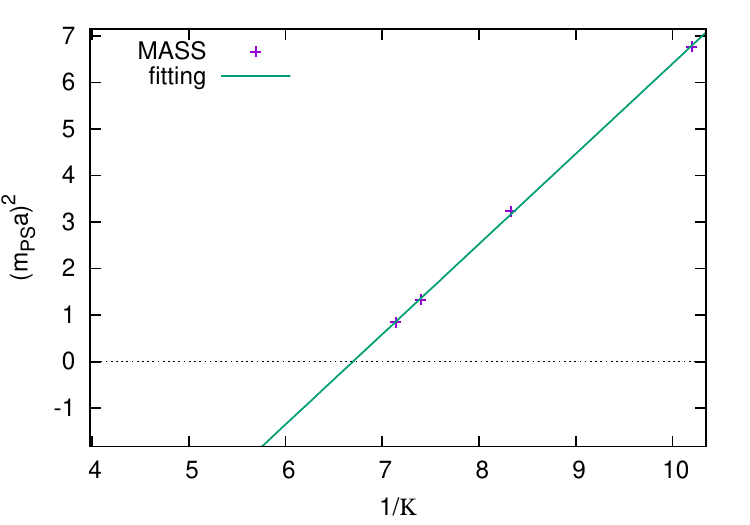}
   \end{center}
   \caption{$(m_{\mathrm{PS}}a)^2$ as a function of $1/K$ at $\beta=5.992$}
   \label{fig:pion_zeropoint}
\end{figure}


To study finite-size effects in this result at $N_t=6$, we also perform simulations on $32^3 \times 6$ and $36^3 \times 6$ lattices.
The results of $V_{\mathrm{eff}}(|\Omega|)$ on the $32^3 \times 6$ lattice are shown in Fig.~\ref{fig:32x6veff_lo}.
We find that the double-well shape becomes milder as we increase $K$.
The height of the potential barrier $\Delta V_{\mathrm{eff}}$ is shown in Fig.~\ref{fig:32x6dveff_lo}.
$\Delta V_{\mathrm{eff}}$ decreases towards zero at $K \sim 0.12$.
Fitting the last seven data points by a linear function, we obtain $K_{ct}=0.1286(40)$,
which is roughly the same as the $K_{ct}$ obtained on the $24^3\times6$ lattice within the statistical error.
The effective potential $V_{\mathrm{eff}}$ on the $36^3 \times 6$ lattice is shown in Fig.~\ref{fig:36x6veff_lo}. 
The figure shows that the double-well shape of $V_{\mathrm{eff}}$ remains up to our largest $K$.
To check reliability of these results, we study the overlap problem in the calculation of $V_{\mathrm{eff}}$.
As discussed in Appendix~\ref{sec:multipoint}, when the number of original configurations 
is not large enough around the peaks of the target histogram,
the results obtained using the reweighting method are not reliable. 
In general, the overlap problem is more likely to occur as the volume increases.

In Figs.~\ref{fig:24t6_allplot} and~\ref{fig:32t6_allplot} we plot the distribution of $(P,|\Omega|)$ from the original configurations at $K=0$ obtained on the $24^3\times6$ and $32^3\times6$ lattices, respectively.
We change color depending on the $\beta$ of the original configurations used in the multi-point histogram method. 
The circles denote the two peak positions of the reweighted histogram at the transition point at various $K$.
We see that the peak positions remain within the region of the original distribution up to the largest $K$ we study.
We thus conclude that the overlap problem does not contaminate the results on the $24^3\times6$ and $32^3\times6$ lattices.

The distribution of $(P,|\Omega|)$ on the largest $36^3\times6$ lattice is shown in Fig.~\ref{fig:36t6_allplot}. 
The two peak positions of the histogram for $K=0.0,$ 0.107 and 0.119 are marked by the circles.
We note that the peak positions for $K=0.119$ are clearly outside of the distribution,
meaning that the overlap problem occurs around there.
We conclude that the result on the $36^3\times6$ lattice is not reliable at $K \approx 0.119$. 
This explains the strange behavior of the histogram in Fig.~\ref{fig:32x6veff_lo} at $K=0.119$. 
To obtain a reliable $V_{\mathrm{eff}}$ on this lattice, 
we would need full QCD simulations or quenched simulations with an external source term of the Polyakov loop~\cite{kiyohara19}.

\subsection{Results for $N_t=8$}
\label{sec:kappacLO8}

The results for $V_{\mathrm{eff}}(|\Omega|)$ obtained on the $24^3\times8$ lattice are shown in Fig.~\ref{fig:24x8veff_lo}.
We find that the double-well shape of $V_{\mathrm{eff}}$ is quite stable up to $K=0.180$. 
At the same time, from the $(P,|\Omega|)$ distribution shown in Fig.~\ref{fig:24t8_allplot}, 
we find that these peak positions of the histogram for $K=0.0,$ 0.15 and 0.18 are in the area where the number of configurations is enough. 
Therefore, the results from the $24^3\times8$ lattice do not suffer from the overlap problem. 
From the lowest order study, we obtain $K_{ct} > 0.18$ on the $24^3\times8$ lattice.

Because this value of $K_{ct}$ is quite large, we examine the location of the chiral limit.
We perform zero-temperature simulations of two-flavor full QCD at $\beta=5.992$, 
which corresponds approximately to the transition point at these $K$'s determined with the reweighting study using quenched configurations.
The details of the study on the zero-temperature lattice are given in Sec.~\ref{sec:mass}.
In Fig.~\ref{fig:pion_zeropoint}, we plot the results of the pseudo-scalar meson mass squared $(m_{\rm PS} a)^2$ by crosses as a function of $1/K$. 
Because they fall on a straight line we perform a linear fit to obtain the chiral limit around $K \approx 0.15$. 

This implies that the hopping parameter expansion is not applicable at $K \simge 0.15$.
The large value of $K_{ct}$ for $N_t=8$ is obtained by an analysis with the LO hopping parameter expansion.
The problem must be due to the truncation error of the hopping parameter expansion. 
We discuss this issue in~Sec.~\ref{sec:kappacNLO}.

\subsection{Critical point from the leading order hopping parameter expansion}
\label{sec:kappacLO}

\begin{table}[tb]
 \begin{center}
  \caption{Summary of $K_{ct}$ and $K_{ct, \mathrm{eff}}$ for two-flavor QCD using the leading order (LO) or up to the next-to-leading order (NLO) terms of $\ln \det M$. $K_{ct}$ on the $24^3 \times 4$ lattice was obtained in Ref.~\cite{Saito1}.}
    \begin{tabular}{c|c|c|c}
   \hline
   lattice  & $K_{ct}$ up to LO  & $K_{ct}$ up to NLO & $K_{ct,\mathrm{eff}}$ with effective NLO \\ \hline
   $24^3 \times 4$ & 0.0658(3)$(^{+4}_{-11})$  &  -   & 0.0640(10) \\
   $24^3 \times 6$ & 0.1359(30)                   & 0.1202(19)   & 0.1205(23) \\
   $32^3 \times 6$ & 0.1286(40)                   & -  & - \\
   $24^3 \times 8$ & $>$ \ 0.18                   & -  & - \\ \hline
  \end{tabular}
 \label{tab:kct}
 \end{center}
\end{table}

Our results for $K_{ct}$ in two-flavor QCD using the LO terms of $\ln \det M$ are summarized in Table~\ref{tab:kct}. 
Results with terms up to NLO are discussed in the next section.
If the $N_t=4$ and 6 lattices are both in the asymptotic scaling region, we would expect that $K_{ct}$ for $N_t =6$ is $3/2$ times larger than that for $N_t=4$ because $K \sim (m_q a)^{-1} = N_t T_c/m_q$ in the limit $m_q a \ll 1$. 
However, $K_{ct}$ for $N_t =6$ turns out to be about twice as large as that for $N_t=4$.
Although the central values of $K_{ct}$ for $N_t =6$ may indicate a decrease of $K_{ct}$ with increasing spatial volume, our data are not yet sufficient to make a large volume extrapolation.
The spatial volume dependence will be studied in more detail in a separate paper~\cite{kiyohara19}.

\section{Influence of next-to-leading order terms}
\label{sec:kappacNLO}

We now study the influence of the NLO terms of the hopping parameter expansion in the reweighting factor $\ln \det M$ given by Eq.~(\ref{eq:hpe_nlo}). 
As discussed in~Sec.~\ref{sec:heavyQregion}, we disregard the effects of the six-step Wilson loops in this study.
We thus concentrate on how the $O(K^{N_t+2})$ bent Polyakov loops $\Omega_n$ with $N_t+2$ steps (as shown in~Fig.~\ref{fig:bendedpoly} for the case of~$N_t=6$) affect the phase structure computed with only the LO Polyakov loop $\Omega$.

\begin{figure}[tb]
   \begin{minipage}{0.45\hsize}
    \begin{center}
     \includegraphics[width=7.5cm]{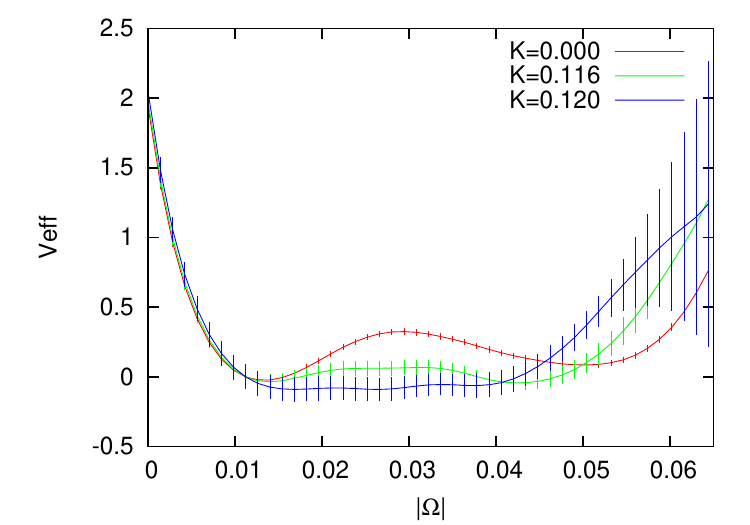}
    \end{center}
    \caption{$V_{\mathrm{eff}}(|\Omega|)$ obtained by the hopping parameter expansion up to the next to leading order on a $24^3\times6$ lattice.}
    \label{fig:24x6veff_nlo}
   \end{minipage}
   \hspace{3mm}
   \begin{minipage}{0.45\hsize}
    \begin{center}
     \includegraphics[width=7.5cm]{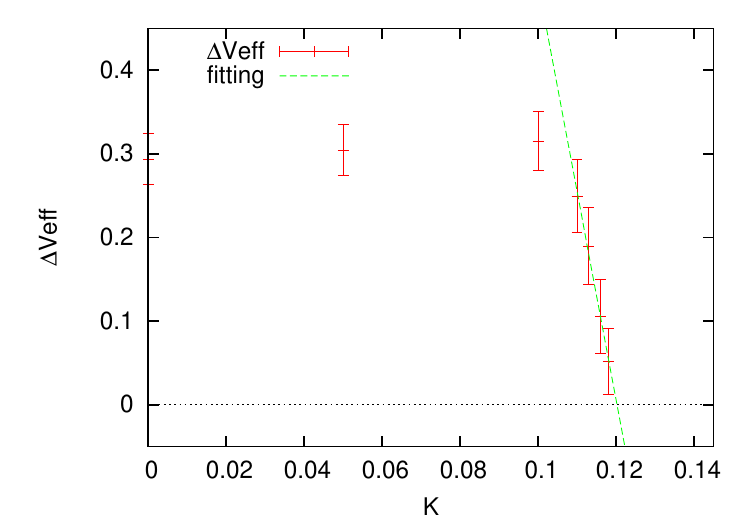}
    \end{center}
    \caption{$\Delta V_{\mathrm{eff}}$ as function of $K$ obtained by the hopping parameter expansion 
    up to the next to leading order on a $24^3\times6$ lattice.}
    \label{fig:24x6dveff_nlo}
   \end{minipage}
  \end{figure}

The NLO results for $V_{\mathrm{eff}}(|\Omega|)$ and  $\Delta V_{\mathrm{eff}}$ obtained on the $24^3\times6$ lattice are shown in~Figs.~\ref{fig:24x6veff_nlo} and~\ref{fig:24x6dveff_nlo}.
The critical point extracted with a linear fit using the four smallest data points of $\Delta V_{\mathrm{eff}}$ turns out to be $K_{ct}=0.1202(19)$. 

\subsection{Effective NLO method}
\label{sec:keff}

\begin{figure}[tb]
    \begin{center}
      \includegraphics[width=7.5cm]{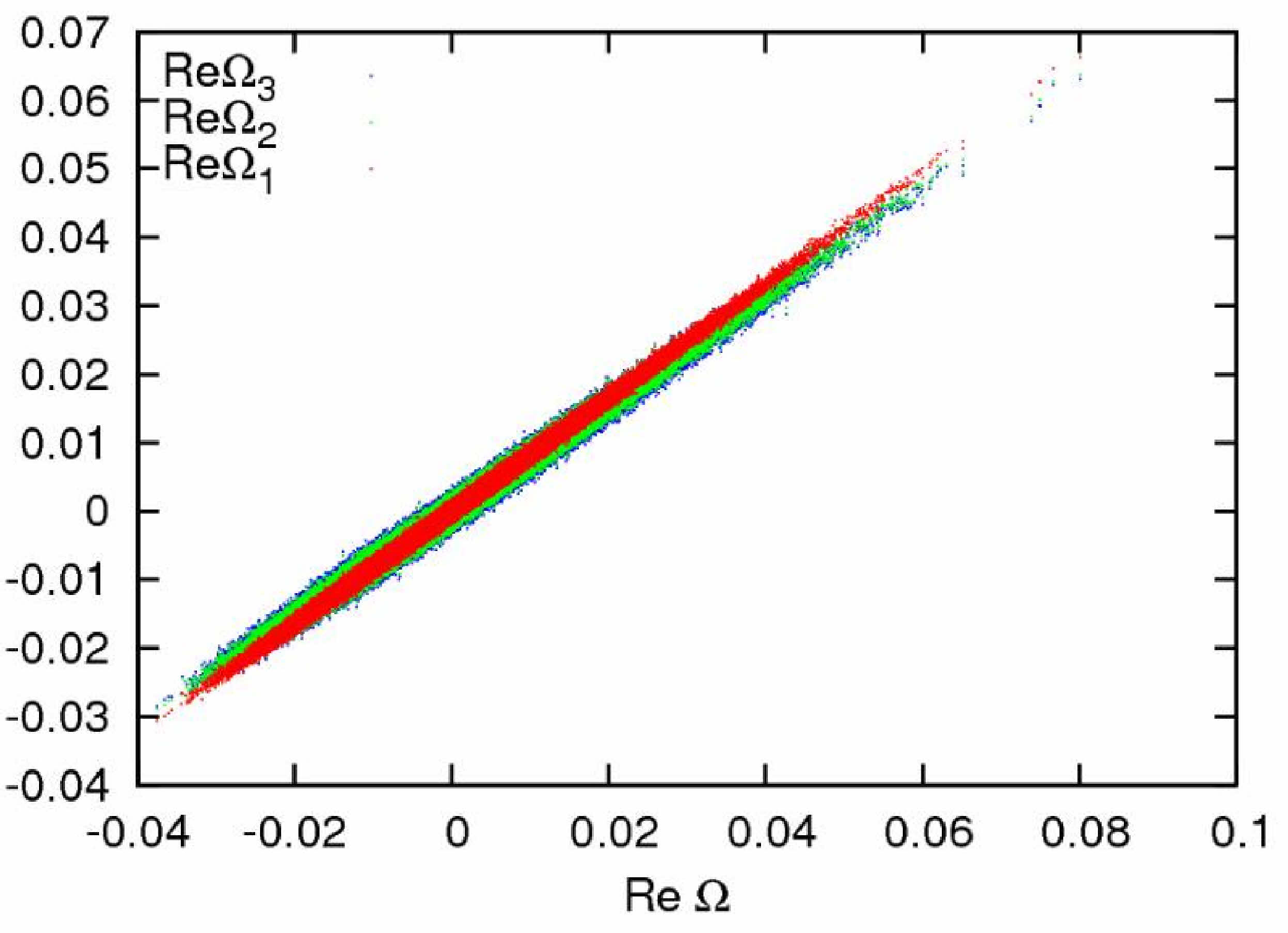}
    \end{center}
    \caption{Distribution of $(\mathrm{Re}\Omega,\mathrm{Re}\Omega_n)$ in quenched QCD on the $24^3\times6$ lattice near the transition point. Red, green, and blue dots are for $n=1$, 2, and 3, respectively.}
    \label{fig:24x6ratio}
   \end{figure}

Before proceeding, let us discuss a method (introduced in~Ref.~\cite{Saito2}) to effectively incorporate NLO terms into the LO calculation of $V_{\mathrm{eff}}$ and $K_{ct}$.
The method is based on the strong linear correlation between $\Omega$ and $\Omega_n$ observed on $N_t=4$ lattices~\cite{Saito2}.
In Fig.~\ref{fig:24x6ratio}, we show the distribution of $(\mathrm{Re} \Omega,\mathrm{Re} \Omega_n)$ measured on each configuration of the $24^3 \times 6$ lattice near the transition point.
Red, green, and blue dots are for $\mathrm{Re} \Omega_1$, $\mathrm{Re} \Omega_2$, and $\mathrm{Re} \Omega_3$, respectively.
We find that $\mathrm{Re} \Omega$ and $\mathrm{Re} \Omega_n$ on the $N_t=6$ lattice also have strong linear correlation with each other.

  \begin{table}[tb]
   \begin{center}
    \caption{The ratio of the Polyakov loops $c_n=\langle \mathrm{Re}\Omega_{n}/\mathrm{Re}\Omega \rangle$ 
    and $C_\Omega$ defined by Eq.~(\ref{eq:COmega}).
    }
    \begin{tabular}{c|ccccc}
     \hline
     lattice      & $\beta$ & $c_1$ & $c_2$ & $c_3$ & $C_\Omega$
     \\ \hline
     $24^3 \times 4$ & 5.6850 & 0.8108(8) & 0.7772(10) & -- & 7.196(6)
     \\
     $24^3 \times 6$ & 5.8750 & 0.8314(4) & 0.8019(6) & 0.7985(7) & 12.195(5)
     \\ \hline
    \end{tabular}
    \label{tab:ratio}
   \end{center}
  \end{table}

With the strong linear correlation, we may introduce an approximation that
\begin{equation}
 \mathrm{Re} \hat\Omega_{n} \approx  c_n \, \mathrm{Re} \hat\Omega, \ \ \ 
n=1,2, \cdots, N_t/2,
\end{equation}
where
$c_n =\langle {\rm Re}\Omega_n / {\rm Re}\Omega \rangle$.
With this approximation, the $O(K^{N_t})$ and $O(K^{N_t+2})$ terms in~Eq.~(\ref{eq:hpe_nlo}) read 
\begin{equation}
12\times 2^{N_t}N_s^3 K^{N_t} \left( 1 + C_\Omega \, N_t K^2 \right) \mathrm{Re} \hat\Omega,
\end{equation}
with
\begin{equation}
C_\Omega = 6 \sum_{n=1}^{N_t/2-1} c_n +3 \, c_{N_t/2}.
\label{eq:COmega}
\end{equation}
This means that the NLO effects can be reproduced by a shift of the hopping parameter in front of $\mathrm{Re} \Omega$ in the LO calculation.
The results for $c_n = \langle \mathrm{Re} \Omega_{n} / \mathrm{Re} \Omega \rangle$ are given in~Table~\ref{tab:ratio} for $24^3\times4$ and $24^3\times6$ lattices.
The $\beta$ dependence in $c_n$ is found to be small in the range we investigated. 
Thus, the values of $c_n$ and $C_{\Omega}$ in~Table~\ref{tab:ratio} are typical values obtained at one $\beta$ and are used in the following calculations.

In particular, the critical point $K_{ct,\mathrm{LO}}$ obtained by the LO calculation can be effectively translated to the critical point $K_{ct,\mathrm{eff}}$ to NLO accuracy by 
\begin{equation}
 K_{ct,\mathrm{eff}} 
  \left( 1 + C_\Omega \, N_tK^2_{ct,\mathrm{eff}} \right)^{1/N_t} 
 = K_{ct,\mathrm{LO}}
\label{eq:keff}
\end{equation}
Solving this relation with the $K_{ct,\mathrm{LO}}$ obtained on the $24^3 \times 6$ lattice, we find $K_{ct,\mathrm{eff}}=0.1205(23)$.
This is well consistent with $K_{ct}=0.1202(19)$ obtained with the direct NLO calculation.
We thus find that the effective NLO method works well.
The method is useful to avoid repeating similar analyses, e.g., for various numbers of flavors.

We also improve the calculation of $K_{ct,\mathrm{eff}}$ for $N_t=4$ in~Ref.~\cite{Saito2}, 
by taking the slight $n$ dependence of $c_n$ into account.
We obtain $K_{ct,\mathrm{eff}}=0.0640(10)$ on the $24^3 \times 4$ lattice. 

\subsection{Critical point with NLO contributions}

Our results for $K_{ct}$ are summarized in~Table~\ref{tab:kct}.
On the $24^3 \times 4$ lattice, we obtain $K_{ct,\mathrm{eff}}=0.0640(10)$ using the effective NLO method.
This is 3\% smaller than $K_{ct,\mathrm{LO}}=0.0658(3)(^{+4}_{-11})$ computed with only the LO contribution, but the difference, i.e., the truncation error of the hopping parameter expansion of $\ln \det M$, is small in comparison with the statistical errors. 

On the other hand, the truncation error turned out to be not negligible for $N_t=6$.
With NLO contributions, we find $K_{ct}=0.1202(19)$, which is about 10\% smaller than the LO value 0.1359(30). 
As expected, the NLO terms reduce $K_{ct}$.
However, the reduction is not sufficient to achieve the $K_{ct} \approx 0.1$ 
expected from the naive scaling with the $N_t=4$ result.

We reserve a study of NLO effects on $N_t=8$ lattices for future work.
The effective method, with which the calculation of bent Polyakov loops can in part be reduced, may be useful on a lattice with larger $N_t$ in which wider variety of bent Polyakov loops have to be calculated.

\section{Meson mass at the critical point}
\label{sec:mass}

To clarify the physical implication of the values for the critical hopping parameter $K_{ct}$ calculated in the previous sections, we calculate the pseudo-scalar meson mass $m_{\mathrm{PS}}$ corresponding to $K_{ct}$ by performing additional zero-temperature simulations.
In this study, we perform two simulations:
a direct two-flavor full QCD simulation adopting the same combination of gauge and quark actions and adjusting the simulation parameters $(\beta, K)$ to the critical point obtained in the finite-temperature study;
and, a quenched QCD simulation combined with the reweighting method, as adopted in the determination of $K_{ct}$ at finite temperatures.
In the latter approach, the LO hopping parameter expansion is adopted, though the influence of $\Omega$ is quite small 
because $\Omega \approx 0$ on zero-temperature configurations.
As discussed in~Sec.~\ref{sec:heavyQregion}, the effect of the plaquette term can be absorbed by the shift $\beta \rightarrow \beta^*$, with $\beta^*$ given by Eq.~(\ref{eq:betastar}).

Our simulation parameters are summarized in Table~\ref{tab:crpt_nf2}.
The combination $(\beta,K)$ is used in the two-flavor full QCD simulations, while $\beta^*$ is used in the quenched QCD simulations.
The hopping parameter $K$ is adjusted to the critical point $K_{ct}$ obtained by finite-temperature simulations 
on $24^3 \times 4$ and $24^3 \times 6$ lattices using the LO and NLO reweighting factors.
In both full and quenched simulations, we generate configurations using the hybrid Monte Carlo algorithm on $16^3 \times 32$ lattices. 
The number of configurations is 52 for each simulation point. 
The meson correlation functions are computed every 10 trajectories after thermalization using a point quark source.
We also perform additional simulations at a few points of $K$ around the simulation points given in Table~\ref{tab:crpt_nf2} 
to evaluate $d(m_{\mathrm{PS}} a)/dK$.
The information about the derivative is used to estimate the error of 
$m_{\mathrm{PS}}$ due to the ambiguity of $K_{ct}$, and also to calculate $m_{\mathrm{PS}}$ at $K_{ct,\mathrm{eff}}$ which is slightly different from the simulation point $K_{ct}$.

\begin{table}[tb]
\begin{center}
\caption{Simulation parameters of zero-temperature simulations on $16^3 \times 32$ lattices.
The combination $(\beta,K)$ is used in the two-flavor full QCD simulations, while $\beta^*$ is used in the quenched QCD simulations for reweighting studies.
$K$ is adjusted to the critical point $K_{ct}$ obtained by finite-temperature simulations 
on $24^3 \times 4$ and $24^3 \times 6$ lattices using the LO or NLO reweighting factor.}
  \begin{tabular}{c|ccc|ccc}
   \hline
   \multicolumn{1}{c|}{$T>0$ lattice } & \multicolumn{3}{c|}{up to LO}   & \multicolumn{3}{c}{up to NLO}  \\ \hline
       & $K$  & $\beta$  & $\beta^*$ & $K$ & $\beta$ & $\beta^*$  \\ \hline
   $24^3 \times 4$ & 0.0658  & 5.680 & 5.682  &  0.0639 & 5.680 & 5.682 \\
   $24^3 \times 6$ & 0.1359  & 5.840 & 5.873  &  0.1202 & 5.852 & 5.872  \\ \hline
  \end{tabular}
  \label{tab:crpt_nf2}
\end{center}
\end{table}

\begin{figure}[tb]
  \begin{center}
\includegraphics[width=7cm]{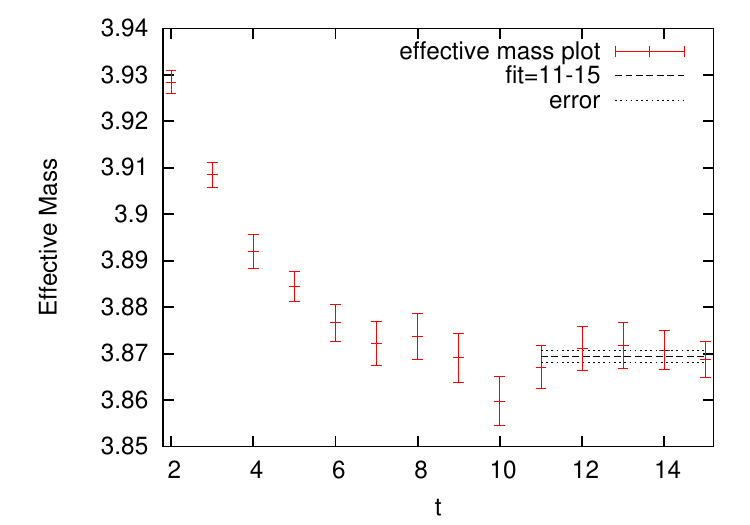}
 \hspace{3mm}
\includegraphics[width=7cm]{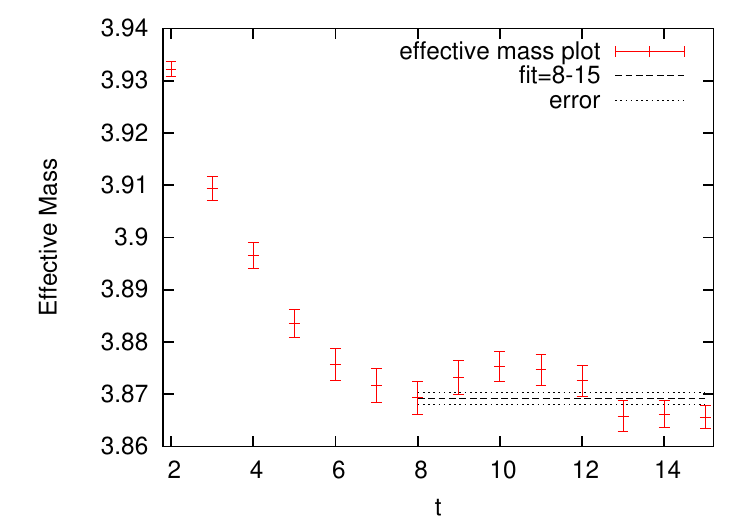}
\includegraphics[width=7cm]{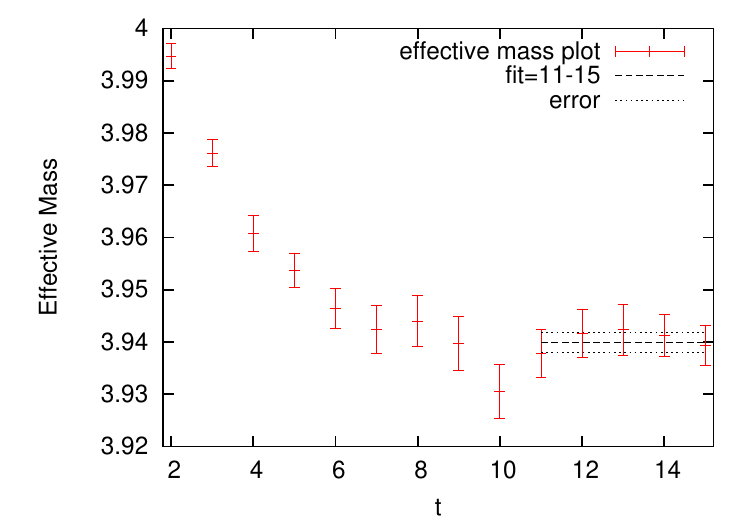}
 \hspace{3mm}
\includegraphics[width=7cm]{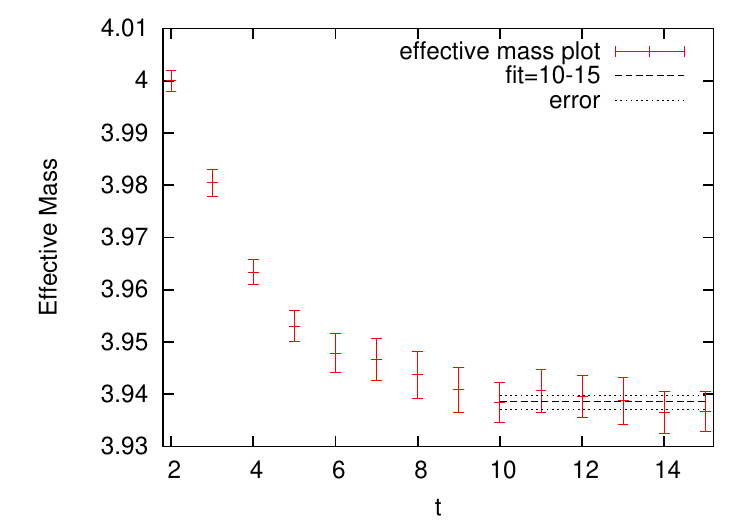}
\includegraphics[width=7cm]{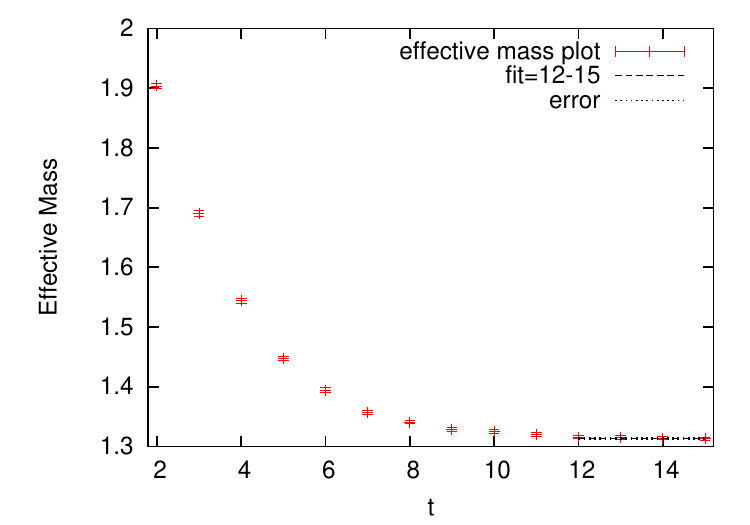}
 \hspace{3mm}
\includegraphics[width=7cm]{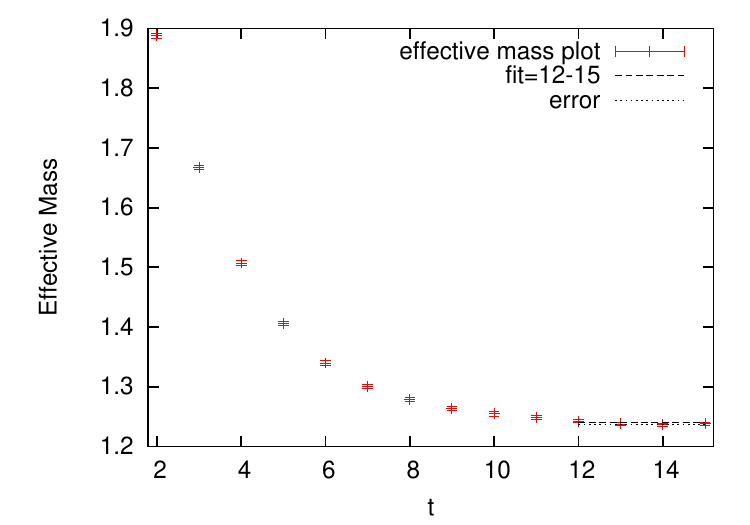}
\includegraphics[width=7cm]{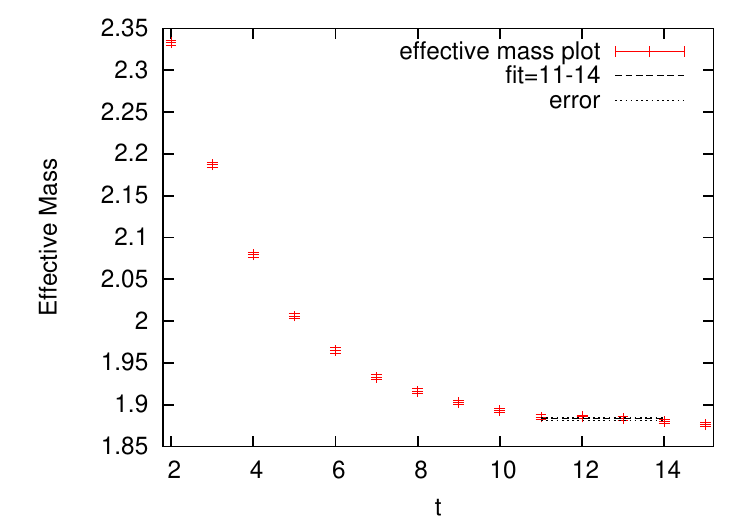}
 \hspace{3mm}
\includegraphics[width=7cm]{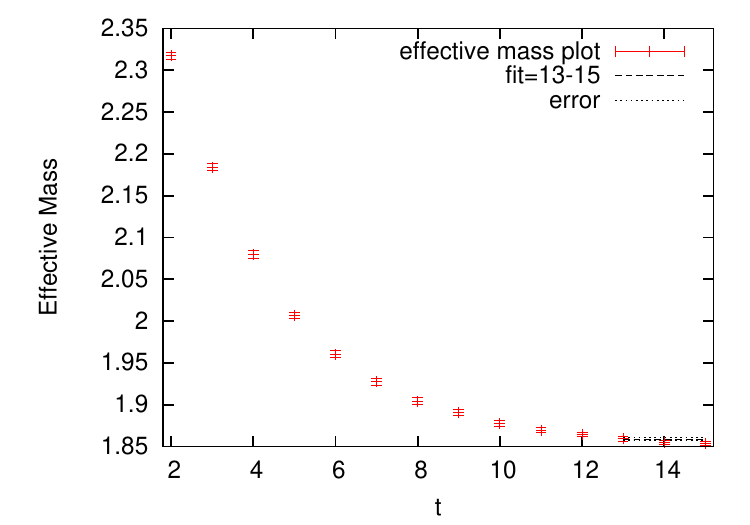}
  \end{center}
\caption{Effective pseudo-scalar meson mass at the critical point.
The left and right panels are obtained by the quenched simulation with reweighting and by the two-flavor full QCD simulation, respectively.
From top to bottom, 
the simulation point is tuned to the critical point of the $24^3 \times 4$ lattice determined by the LO calculation, 
the $24^3 \times 4$ lattice by the NLO calculation, 
the $24^3 \times 6$ lattice by the LO calculation, and 
the $24^3 \times 6$ lattice by the NLO calculation,
respectively.
The horizontal solid lines are the results for $m_{\mathrm{PS}}a$ obtained by the fit using Eq.~(\ref{eq:massfit}), with the fit range shown by the extent of the horizontal lines.
}
  \label{fig:meff}
\end{figure}


To evaluate $m_{\mathrm{PS}}a$, we fit the pseudo-scalar meson correlation function $\langle \pi (t) \pi(0) \rangle$ 
using the following fit function:
\begin{eqnarray}
\left\langle \pi(t)\pi(0) \right\rangle 
\approx C \left[ e^{-m_{\mathrm{PS}}a t} +  e^{-m_{\mathrm{PS}}a (N_t -t)} \right]
= C e^{-m_{\mathrm{PS}}a N_t/2} \cosh \left[ m_{\mathrm{PS}}a (t-N_t/2) \right], 
\label{eq:massfit}
\end{eqnarray}
where $m_{\mathrm{PS}}a$ and $C$ are the fit parameters.
The fit range is decided by considering the effective mass defined by
\begin{equation}
 m_{\rm eff}(t) a = \cosh^{-1}  \left[\frac{\langle \pi (t+1) \pi(0) \rangle + \langle \pi (t-1) \pi(0) \rangle}{
2\langle \pi (t) \pi(0) \rangle}  \right].
\end{equation}
This effective mass is constant with $m_{\mathrm{PS}}a$ when the correlation function is well approximated by Eq.~(\ref{eq:massfit}).
We choose the fit range where the effective mass exhibits a plateau.

Our results for the effective mass are shown in Fig.~\ref{fig:meff} as function of $t$. 
The left panels are obtained by the quenched simulation with reweighting, and
the right panels are obtained by two-flavor full QCD simulations.
The top two plots are obtained at $K_{ct}$ of the LO for the $24^3 \times 4$ lattice.
The two plots in the second row are obtained at $K_{ct}$ of the NLO for the $24^3 \times 4$ lattice, 
those in the third row are at $K_{ct}$ of the LO for the $24^3 \times 6$ lattice,
and those in the bottom row are at $K_{ct}$ of the NLO for the $24^3 \times 6$ lattice.
The horizontal solid lines represent the fit range and, simultaneously, the results for $m_{\mathrm{PS}}a$ with their statistical errors shown by the dashed lines.

The results for the pseudo-scalar meson mass are summarized in~Table \ref{tab:mass_nf2}.
The errors for $m_{\mathrm{PS}}a$ in~Table~\ref{tab:mass_nf2} represent the statistical errors from the hadron mass fit.
On the other hand, in the errors for $m_{\mathrm{PS}}/T_c = N_t \times (m_{\mathrm{PS}}a)$, we include those propagated from the error of $K_{ct}$ using $d(m_{\mathrm{PS}} a)/dK$ obtained by additional simulations near $K_{ct}$.

Comparing the results of the two-flavor full QCD simulation with those of the quenched simulation with reweighting, 
we find that the results for $N_t=4$ are well consistent with each other. 
Although the results for $m_{\mathrm{PS}}a$ show slight deviation for the case of $N_t=6$, the deviation is much smaller than the error propagated from the error of $K_{ct}$, as shown by $m_{\mathrm{PS}}/T_c$.
Recall that the six-step Wilson loops are neglected in the reweighting calculation, 
while the quark determinant is fully included in the full QCD simulation.
Hence, the consistency of reweighting and full QCD results means that the truncation error of the hopping parameter expansion is small in the calculation of $m_{\mathrm{PS}}$ at the critical points on $N_t=4$ and 6 lattices.
Because the computational cost of the quenched simulation is much smaller than that of the full QCD simulation, 
we may compute other hadron masses at $K_{ct}$ using the quenched simulation with reweighting.

On the other hand, the truncation error in the calculation of 
$K_{ct}$ is large for the $N_t=6$ lattice as discussed in~Sec.~\ref{sec:kappacNLO}.
In accordance with this, the value of $m_{\mathrm{PS}}/T_c$ at the LO $K_{ct}$ is about $1.4$ times larger than that at the NLO $K_{ct}$ for $N_t=6$.
If we assume that the systematic error from the truncation of the hopping parameter expansion is roughly the same as the difference between the LO and NLO results, we obtain $m_{\mathrm{PS}}/T_c = 11.15(42)(372)$ for $N_t=6$.
This is smaller than the result $m_{\mathrm{PS}}/T_c = 15.73(14)(26)$ for $N_t=4$.
Hence, our results suggest that the critical pseudo-scalar meson mass measured on the $N_t=6$ lattice is smaller than that on the $N_t=4$ lattice. 

\begin{table}[tb]
  \caption{Pseudo-scalar meson mass at the critical point of two-flavor QCD, 
  computed by quenched simulation with reweighting and by two-flavor full QCD simulation.}
 \begin{center}
  \begin{tabular}{c|cc|cc}
   \hline
   \multicolumn{1}{c|}{$T>0$ lattice} & \multicolumn{2}{c|}{$m_{\mathrm{PS}}a$}    & \multicolumn{2}{c}{$m_{\mathrm{PS}}/T_c$}  \\ \hline
   $24^3 \times 4$  & LO  & NLO & LO & NLO  \\ \hline
   reweighting & 3.8694(12) & 3.9353(20) & 15.47(14) & 15.74(14) \\
   full QCD  & 3.8692(11)   & 3.9342(14) & 15.47(14) & 15.73(14) \\ \hline
   $24^3 \times 6$  & LO & NLO  & LO  & NLO  \\ \hline
   reweighting &  1.3141(22)  & 1.8831(12) & 7.88(69) & 11.29(40) \\ 
   full QCD  & 1.2394(18)  & 1.8590(19) & 7.43(78)  & 11.15(42) \\ \hline
  \end{tabular}
  \label{tab:mass_nf2}
 \end{center}
\end{table}

\section{Critical line in 2+1 flavor QCD}
\label{sec:2+1flavor}


We extend the analysis of the critical point to $2+1$ flavor QCD.
Up to the LO of the hopping parameter expansion, the quark determinant term is given by
\begin{equation}
 \ln \left[ \frac{(\mathrm{det}M(K_{ud}))^2 \mathrm{det}M(K_{s})
      }{(\mathrm{det}M(0))^3 }\right] =
 288N_{\mathrm{site}}(2K_{ct,ud}^4+K_{s}^4)\hat{P} + 12\times 2^{N_t}N_s^3
 (2K_{ud}^{N_t}+K_{s}^{N_t}) \mathrm{Re} \hat{\Omega},
\label{eq:2+1detm}
\end{equation}
where $K_{ud}$ and $K_s$ are the hopping parameters of the degenerate up and down quarks and of the strange quark, respectively. 
The case of two-flavor QCD is reproduced by setting $K_s=0$ in~Eq.~(\ref{eq:2+1detm}).
As discussed in~Sec.~\ref{sec:heavyQregion}, the effects of the plaquette term can be absorbed by a shift of the gauge coupling $\beta$. 
We then find that the reweighting factor for $2+1$ flavor QCD is obtained by replacing $2K^{N_t}$ in two-flavor QCD with $2K_{ud}^{N_t}+K_{s}^{N_t}$.
In particular, the critical line in the $(K_{ud},K_s)$ plane of $N_{\rm f}=2+1$ QCD
and the LO critical point $K_{ct,\mathrm{LO}}$ of $N_{\rm f}=2$ QCD are related as \cite{Saito1}
\begin{equation}
 2K_{ct,ud}^{N_t}+K_{ct,s}^{N_t} = 2K_{ct,\mathrm{LO}}^{N_t}.
\label{eq:2+1clLO}
\end{equation}


We can easily extend this relation in the LO to that in the NLO applying the effective NLO method discussed in~Sec.~\ref{sec:keff}. 
By substituting Eq.~(\ref{eq:keff}) into Eq.~(\ref{eq:2+1clLO}), we obtain an effective NLO relation for the critical line,
\begin{align}
2 K_{ct,ud}^{N_t}  \left( 1 + C_\Omega \, N_tK^2_{ct,ud} \right) 
+ K_{ct,s}^{N_t}  \left( 1 + C_\Omega \, N_tK^2_{ct,s} \right)
=2 K_{ct,\mathrm{LO}}^{N_t}  ,
\label{eq:2+1clNLO}
\end{align}
with $C_\Omega$ given by Eq.~(\ref{eq:COmega}).

\begin{figure}[tb]
 \begin{minipage}{0.45\hsize}
  \begin{center}
  \vspace{10mm}
   \includegraphics[width=7.5cm]{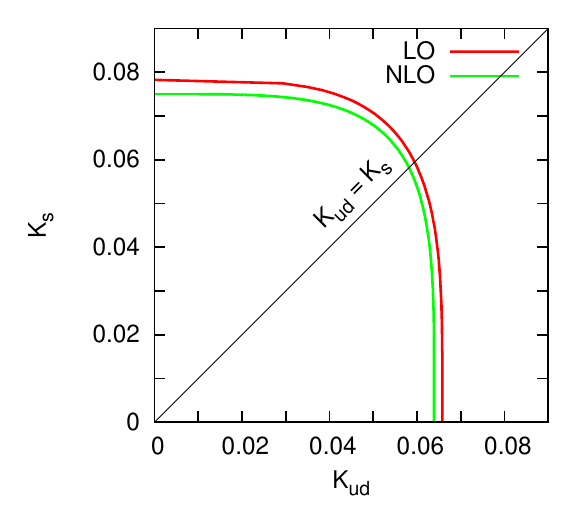}
  \vspace{-10mm}
  \end{center}
  \caption{Critical line of $2+1$ flavor QCD on the $24^3 \times 4$ lattice, determined using the leading order (LO) and the next-to-leading order (NLO) reweighting factors.}
  \label{fig:nt4Kct}
 \end{minipage}
 \hspace{3mm}
 \begin{minipage}{0.45\hsize}
  \begin{center}
  \vspace{10mm}
   \includegraphics[width=7.5cm]{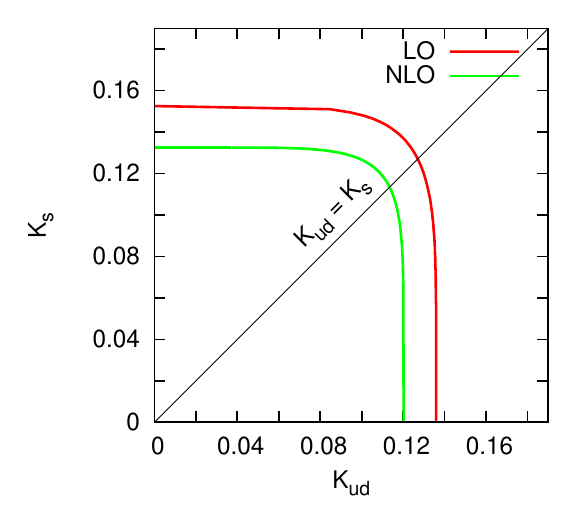}
  \vspace{-10mm}
  \end{center}
  \caption{The same as Fig.~\ref{fig:nt4Kct} but on the $24^3 \times 6$ lattice.}
  \label{fig:nt6Kct}
 \end{minipage}
\end{figure}

\begin{table}[tb]
\caption{The critical $K$ of $N_{\rm f}$ flavor QCD determined by the leading order (LO) and the next-to-leading order (NLO) calculations on the $24^3 \times 4$ and $24^3 \times 6$ lattices.}
\begin{center}
 \begin{tabular}{c|ccc}
 \hline
   $K_{ct}$ & $N_{\rm f} = 1$ & $N_{\rm f} = 2$ & $N_{\rm f} = 3$ \\ \hline
   $24^3 \times 4$, LO  & 0.0783(12) & 0.0658(10) & 0.0595(10) \\ 
   $24^3 \times 4$, NLO & 0.0753(11) & 0.0640(10) & 0.0582(9) \\
   $24^3 \times 6$, LO  & 0.1525(34) & 0.1359(30) & 0.1270(28) \\
   $24^3 \times 6$, NLO & 0.1326(21) & 0.1202(19) & 0.1135(18) \\ \hline
  \end{tabular}
 \label{tab:kctnf}
 \end{center}
\end{table}

We solve Eqs.~(\ref{eq:2+1clLO}) and (\ref{eq:2+1clNLO}) 
using the results for $K_{ct,\mathrm{LO}}$ obtained on $24^3 \times 4$ and $24^3 \times 6$ lattices.
The results for LO and NLO critical lines are shown in Figs.~\ref{fig:nt4Kct} and \ref{fig:nt6Kct} for $N_t=4$ and 6, respectively.
The red and green curves are for the LO and NLO calculations, respectively.
The difference between the two curves shows the amount of truncation error in the hopping parameter expansion.
The results for $K_{ct}$ obtained by the LO and effective NLO calculations for QCD with $N_{\rm f}$ degenerate flavors are summarized in Table \ref{tab:kctnf}
\footnote{
We note that, for $N_{\rm f} = 2$ and $N_t = 6$, the result of $K_{ct}$ using the effective NLO method is $0.2\%$ larger than the direct calculation of $K_{ct,\mathrm{NLO}}$.
In order to match the result of the $N_{\rm f} = 2$ effective NLO method with the result of direct calculation, in Table~\ref{tab:kct} we have replaced the right hand side of Eq.~(\ref{eq:2+1clNLO}),
$2 K_{ct,\mathrm{LO}}^{N_t}$, with
$2 K_{ct,\mathrm{NLO}}^{N_t}(1+ C_\Omega N_t K^2_{ct,\mathrm{NLO}})$
for $N_t = 6$.
}. 
The results for $N_{\rm f} = 1,$ 2 and 3 are the values of $K_{\rm s}$ at $K_{\rm ud}=0$, $K_{\rm ud}$ at $K_{\rm s}=0$ and $K$ when $K_{\rm ud} = K_{\rm s}$ 
on the critical lines in Figs.~\ref{fig:nt4Kct} and \ref{fig:nt6Kct}, respectively.

\begin{figure}[tb]
 \begin{minipage}{0.45\hsize}
  \begin{center}
  \vspace{8mm}
   \includegraphics[width=6.5cm]{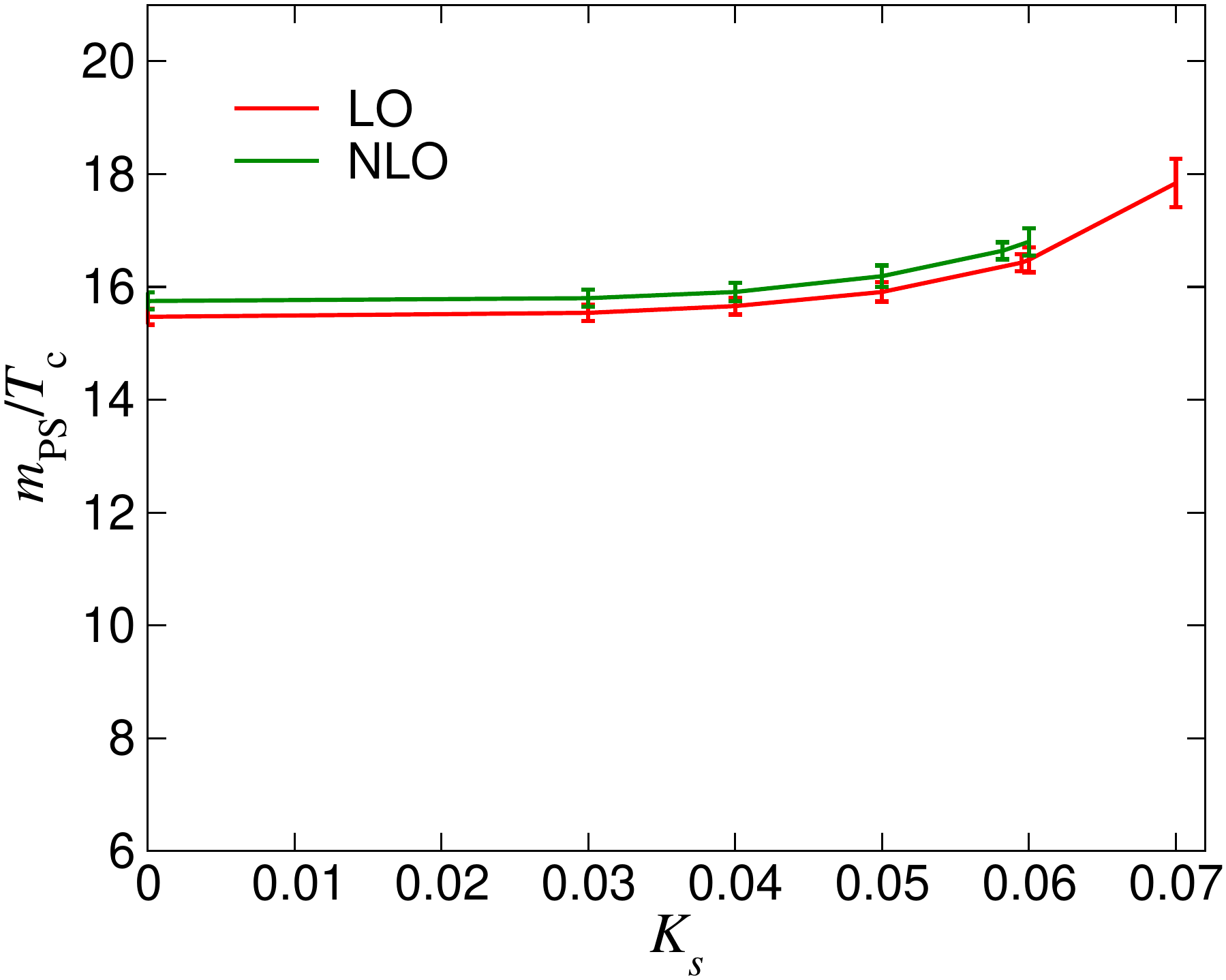}
  \vspace{-8mm}
  \end{center}
  \caption{$m_{\mathrm{PS}}/T_c$ on the critical line for $N_t=4$ 
  as function of $K_{ct,s}$.}
  \label{fig:2+1massNt4}
 \end{minipage}
 \hspace{3mm}
 \begin{minipage}{0.45\hsize}
  \begin{center}
  \vspace{8mm}
   \includegraphics[width=6.5cm]{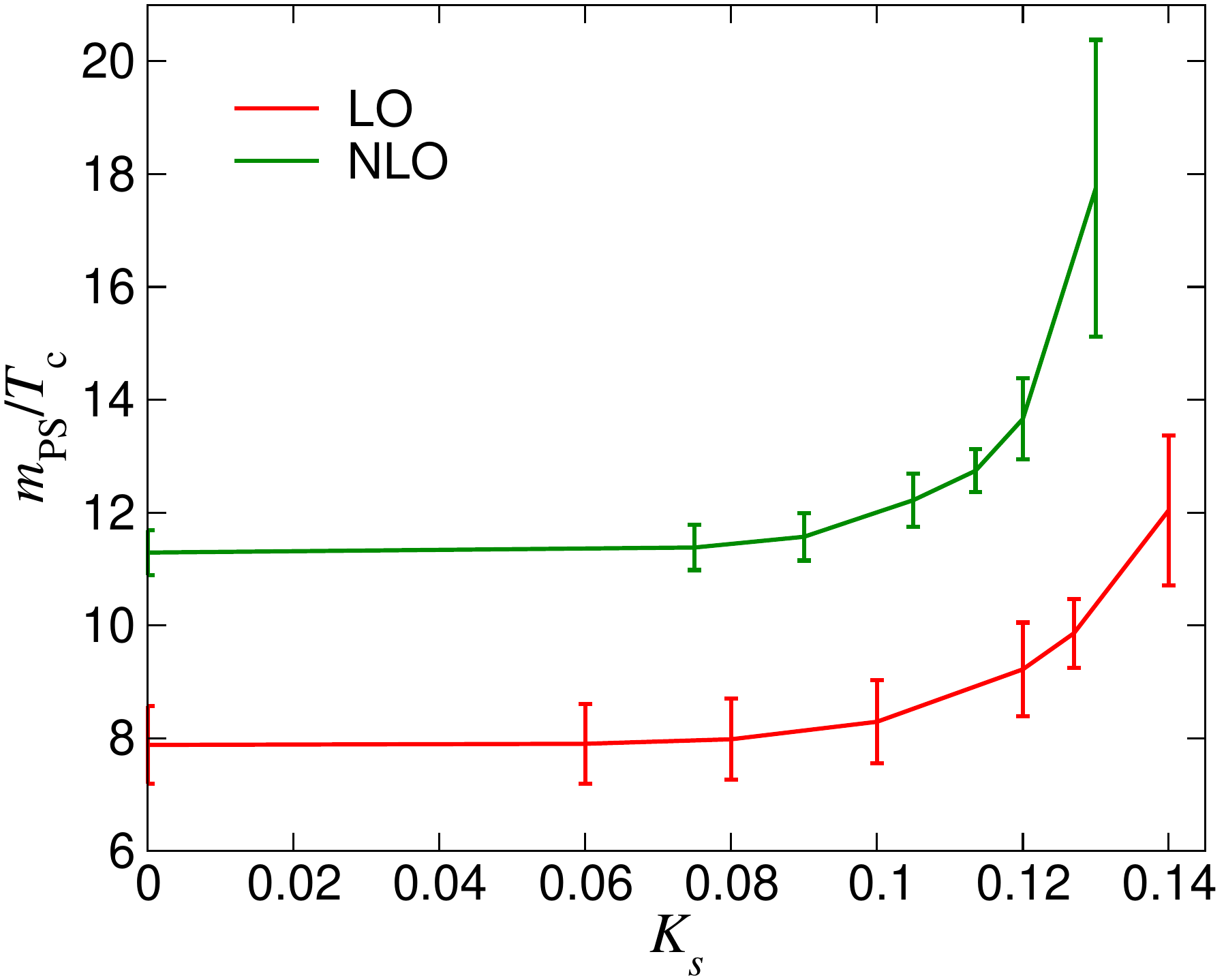}
  \end{center}
  \vspace{-8mm}
  \caption{The same as Fig.~\ref{fig:2+1massNt4} but for $N_t=6$.}
  \label{fig:2+1massNt6}
 \end{minipage}
\end{figure}

Finally, we calculate the pseudo-scalar meson mass on the critical line adopting the reweighting method discussed in Sec.~\ref{sec:mass}.
In the reweighting method, we shift $\beta$ to $\beta^* = \beta+48(2K_{ud}^4 + K_s^4)$ for $2+1$ flavor QCD.
Here, $\beta^*$ should be fine-tuned to the transition point depending on the values of $K_{ct,ud}$ and $K_{ct,s}$.
However, because $m_{\mathrm{PS}}/T_c$ is not so sensitive to $\beta$, 
we adopt $\beta^*=5.680$ and $5.870$ for all $(K_{ct,ud},K_{ct,s})$ on the $N_t=4$ and 6 lattices, respectively.
We compute only the mass of the pseudo-scalar meson that contains two degenerate light quarks and no strange quark.
The results for $m_{\mathrm{PS}}a$ and $m_{\mathrm{PS}}/T_c$ as functions of 
 $(K_{ct,ud}, K_{ct,s})$ are summarized in Table~\ref{tab:2+1mass}. 
We have checked at several simulation points that $m_{\mathrm{PS}}/T_c$ does not change within the error even if $\beta^*$ is fine-tuned to the transition point.
Our results for $m_{\mathrm{PS}}/T_c$ are plotted in Fig.~\ref{fig:2+1massNt4} $(N_t=4)$ and 
Fig.~\ref{fig:2+1massNt6} $(N_t=6)$ as function of $K_s$.
As in the case of two-flavor QCD, we find that $m_{\mathrm{PS}}/T_c$ on the $N_t=6$ lattice is smaller than that on the $N_t=4$ lattice.
We also note that the truncation error of the hopping parameter expansion is large for $N_t=6$. 

\begin{table}[tb]
 \caption{Pseudo-scalar meson mass on the critical line of $2+1$ flavor QCD
 computed by quenched simulation with reweighting.}
 \begin{center}
  \begin{tabular}{cccc}
   \hline
   \multicolumn{4}{c}{critical line for $N_t=4$, up to LO}  \\ \hline
   $K_{ct,s}$ & $K_{ct,ud}$   & $m_{\mathrm{PS}}a$ & $m_{\mathrm{PS}}/T_c$
   \\ \hline
   0.0000       & 0.0658(10)    & 3.8694(12)    & 15.47(14)    \\
   0.0300       & 0.0654(10)    & 3.8849(7)      & 15.54(14)    \\
   0.0400       & 0.0646(11)    & 3.9151(7)      & 15.66(15)    \\
   0.0500       & 0.0629(12)    & 3.9779(7)      & 15.91(17)    \\
   0.0595       & 0.0595(10)    & 4.1085(6)      & 16.43(15)     \\
   0.0600       & 0.0592(14)    & 4.1204(6)      & 16.48(22)    \\
   0.0700       & 0.0510(23)   & 4.4594(5)      & 17.84(43)     \\ \hline
   \hline
   \multicolumn{4}{c}{critical line for $N_t=4$, up to NLO}  \\ \hline
  $K_{ct,s}$ & $K_{ct,ud}$   & $m_{\mathrm{PS}}a$ & $m_{\mathrm{PS}}/T_c$
   \\ \hline
   0.0000     & 0.0640(10)   & 3.9398(19)    & 15.75(15)    \\
   0.0300     & 0.0636(10)   & 3.9512(7)      & 15.80(15)    \\
   0.0400     & 0.0629(11)   & 3.9779(7)      & 15.91(16)    \\
   0.0500     & 0.0611(12)   & 4.0467(7)      & 16.19(19)    \\
   0.0582     & 0.0582(9)    & 4.1598(6)     & 16.64(15)    \\
   0.0600     & 0.0572(14)   & 4.1996(6)      & 16.80(24)    \\ \hline
   \hline
   \multicolumn{4}{c}{critical line for $N_t=6$, up to LO}  \\ \hline
   $K_{ct,s}$ & $K_{ct,ud}$   & $m_{\mathrm{PS}}a$ & $m_{\mathrm{PS}}/T_c$
   \\ \hline
   0.0000     & 0.1359(30)   & 1.3141(22)      & 7.88(69)    \\
   0.0600     & 0.1358(30)   & 1.3179(10)      & 7.90(71)    \\
   0.0800     & 0.1355(30)   & 1.3296(10)      & 7.98(72)    \\
   0.1000     & 0.1341(32)   & 1.3814(10)      & 8.29(74)    \\
   0.1200     & 0.1299(37)   & 1.5369(11)      & 9.22(83)    \\
   0.1270     & 0.1270(28)   & 1.6427(11)      & 9.86(61)    \\
   0.1400     & 0.1168(63)   & 2.0061(12)      & 12.04(133)   \\ \hline
   \hline
   \multicolumn{4}{c}{critical line for $N_t=6$, up to NLO}  \\ \hline
   $K_{ct,s}$ & $K_{ct,ud}$   & $m_{\mathrm{PS}}a$ & $m_{\mathrm{PS}}/T_c$
   \\ \hline
   0.0000     & 0.1202(19)   & 1.8831(12)      & 11.29(40)    \\ 
   0.0750     & 0.1199(20)   & 1.8962(12)      & 11.38(40)    \\
   0.0900     & 0.1190(21)   & 1.9284(12)      & 11.57(42)    \\
   0.1050     & 0.1160(23)   & 2.0368(12)      & 12.22(47)    \\
   0.1135     & 0.1135(18)  & 2.1230(12)      & 12.74(38)    \\
   0.1200     & 0.1091(35)   & 2.2769(12)      & 13.66(72)    \\
   0.1300     & 0.0896(114)  & 2.9578(12)     & 17.75(263)   \\ \hline
  \end{tabular}
 \label{tab:2+1mass}
 \end{center}
\end{table}

\section{Conclusions}
\label{sec:conclusion}

We studied the end point of the first-order phase transition in the heavy quark region of QCD, by computing the effective potential $V_{\rm eff} (|\Omega|)$ defined as the logarithm of the probability distribution function of the absolute value of the Polyakov loop $\Omega$.
We evaluated $V_{\rm eff} (|\Omega|)$ at various values of the hopping parameter $K$ using the reweighting method from the heavy quark limit.
We evaluated the reweighting factor at small $K$ up to the leading $O(K^{N_t})$ as well as the next-to-leading $O(K^{N_t+2})$ of the hopping parameter expansion for the Polyakov-loop-type operators.
To investigate the effective potential in a wide range of $|\Omega|$ covering the hot and cold phases at the first-order phase transition, we combined the data obtained at several simulation points using the multi-point reweighting method.
We found that the clear double-well shape of $V_{\rm eff} (|\Omega|)$ at $K=0$ changes gradually as we increase $K$, and eventually turns into a single-well shape.
We defined the critical point $K_{ct}$ as the point where the double-well shape disappears. 

Our results for $K_{ct}$ for two-flavor QCD are summarized in Table~\ref{tab:kct}.
We found that, although the volume dependence of $K_{ct}$ is small, $K_{ct}$ seems to decrease as the volume increases. 
On our largest lattice, $36^3 \times 6$, however, we could not determine $K_{ct}$ due to the overlap problem.
We reserve a study of the volume dependence for a future work \cite{kiyohara19,Ejiri19}.
Comparing the results for $K_{ct}$ calculated up to the leading and next-to-leading orders of the hopping parameter expansion, 
we found that the truncation error of the hopping parameter expansion is not small at $N_t=6$, while it is negligible at $N_t=4$.
A reason that the convergence of the hopping parameter expansion is worse at $N_t=6$ is that $K_{ct}$ at $N_t=6$ is larger than that at $N_t=4$, as expected from the fact that $1/K$ is approximately proportional to the quark mass times the lattice spacing.
However, by comparing the results for $N_t=4$ and 6, we note that $K_{ct}$ at $N_t=6$ is much larger than that expected from naive scaling with the data at $N_t=4$. 
To make implication of large values of $K_{ct}$ clearer, we also calculated the pseudo-scalar meson mass $m_{\mathrm{PS}}$ at zero-temperature just on the $K_{ct}$ for $N_t=4$ and $N_t=6$.
Although the truncation error of the hopping parameter expansion is large for $N_t=6$, 
our results suggest that the critical pseudo-scalar meson mass measured on $N_t=6$ lattice is smaller 
than that on $N_t=4$ lattice.
This means that the critical quark mass decreases as the lattice spacing decreases.

We also extended the study of two-flavor QCD to $2+1$ flavor QCD and determined the critical line at the boundary of the first-order transition region in the $(K_{ud}, K_s)$ plane.
Our results for the critical line in $2+1$ flavor QCD are summarized in~Figs.~\ref{fig:nt4Kct} and \ref{fig:nt6Kct} for $N_t=4$ and 6, respectively.
The pseudo-scalar meson masses on the critical line are shown in~Figs.~\ref{fig:2+1massNt4} and \ref{fig:2+1massNt6}.
The general characteristics are the same as the two-flavor case.

To extract quantitative physical predictions, we need to reduce the lattice spacing by increasing $N_t$.
However, our test study on an $N_t=8$ lattice suggests that the reweighting from quenched QCD using the low order hopping parameter expansion is not applicable at $N_t > 6$.
Careful treatments including higher order terms of the hopping parameter expansion are required.
One possible approach is to take into account the effects of further higher order terms of the hopping parameter expansion (\ref{eq:tayexp}).
It is not difficult to calculate ${\cal D}_n$ numerically for a given large $n$.
If we use a random noise method, the trace in Eq.~(\ref{eq:derkappa}) can be computed directly without deriving an analytical equation like Eq.~(\ref{eq:hpe_nlo}) for each $n$.
The study by the hopping parameter expansion has an advantage in finite density studies, as shown in Ref.~\cite{Saito2}. 
Thus, it would be interesting to see if $K_{ct}$ can be determined on finer lattices by increasing the number of expansion terms to understand finite density QCD.
We leave these studies to future work.

\vspace{5mm}
\noindent\textbf{Acknowledgments}

This work was supported by in part JSPS KAKENHI (Grant Nos.~JP19H05598, 
JP19K03819, JP19H05146,  JP17K05442, JP15K05041, JP26400251, and JP26287040), 
the Uchida Energy Science Promotion Foundation, 
the HPCI System Research project (Project ID: hp170208, hp190028, hp190036), and Joint Usage/Research Center for Interdisciplinary Large-scale Information Infrastructures in Japan (JHPCN) (Project ID: jh190003, jh190063).
This research used computational resources of SR16000 and BG/Q at the Large Scale Simulation Program of High Energy Accelerator Research Organization (KEK) (No.~16/17-05), 
OCTPUS of the large-scale computation program at the Cybermedia Center, Osaka University,
and ITO of the JHPCN Start-Up Projects at the Research Institute for Information Technology, Kyushu University.
The authors also thank the Yukawa Institute for Theoretical Physics at Kyoto University for the workshop YITP-S-18-01.

\appendix
\section{Multi-point histogram method}
\label{sec:multipoint}

To investigate how the shape of $V_{\mathrm{eff}}$ is changed by varying quantities such as quark mass,
we need to obtain the histogram in a wide range of $|\Omega|$ to some accuracy. 
This is not straightforward with a single simulation because the statistical accuracy of the histogram quickly decreases awy from the peak point.
To confront this issue, we combine information at several simulation points using the multi-point histogram method~\cite{Saito1,Saito2,FS89,Iwami15}. 

Let us consider a double histogram of the plaquette $P$ and Polyakov loop $|\Omega|$,
\begin{eqnarray}
w(P, |\Omega|; \beta, K) 
= \int {\cal D} U \ \delta(P - \hat{P}) \ \delta(|\Omega| - |\hat{\Omega}|) \ 
e^{-S_g}\ \prod_{f=1}^{N_{\rm f}} \det M(K_f) .
\label{eq:dist}
\end{eqnarray}
With the gauge action (\ref{eq:Sg}), and the quark action (\ref{eq:Sq}), which does not depend on $\beta$, 
the coupling parameter $\beta$ can be easily shifted from a simulation point $\beta_i$ to $\beta$ by the reweighting technique as
\begin{eqnarray}
w(P, |\Omega|; \beta_i, K) = e^{6N_{\rm site} (\beta_i-\beta) P}  \,w(P, |\Omega|; \beta, K) .
\end{eqnarray}

Let us first consider the case of combining information obtained at several simulation points $(\beta_i, K)$, with $i=1, 2, \ldots , N_{\rm sim}$, at a fixed $K$.
The naive histogram using all of the configurations, disregarding the differences in the simulation parameters, is given by
\begin{eqnarray}
\sum_{i=1}^{N_{\rm sim}} N_i \, Z^{-1}_i \, w(P, |\Omega|; \beta_i, K) 
= e^{-6N_{\rm site} \beta P}  w(P,|\Omega|; \beta, K)
\sum_{i=1}^{N_{\rm sim}} N_i \, Z^{-1}_i \, e^{6N_{\rm site}  \beta_i P}  , 
\label{eq:sum1}
\end{eqnarray}
where $N_i$ and $Z_i = Z(\beta_i, K)$ are the number of configurations and the partition function at the $i$th simulation point $(\beta_i, K)$, respectively.
From this relation, we obtain a useful expression for the histogram at $(\beta, K)$, 
\begin{eqnarray}
w(P, |\Omega|; \beta, K)= G(P;\beta,\vec\beta) \,
\sum_{i=1}^{N_{\rm sim}} N_i \, Z^{-1}_i \, w(P ,|\Omega|; \beta_i, K) 
\end{eqnarray}
with $\vec\beta=(\beta_1,\cdots,\beta_{N_{\rm sim}})$ and 
\begin{eqnarray}
G(P;\beta,\vec\beta)=\frac{ e^{6N_{\rm site} \beta P}}{
\sum_{i=1}^{N_{\rm sim}} N_i \, Z^{-1}_i e^{6N_{\rm site} \beta_i P} } .
\end{eqnarray}
For notational simplicity, we suppress the dependence on $K$ in $G$.

The partition function $Z(\beta, K)$ is then given by
\begin{eqnarray}
Z(\beta, K)= \sum_{i=1}^{N_{\rm sim}} N_i \int G(P;\beta,\vec\beta) \, Z^{-1}_i \, w(P,|\Omega|; \beta_i, K) \, dP \,d|\Omega| 
=\sum_{i=1}^{N_{\rm sim}} N_i \left\langle G(\hat{P};\beta,\vec\beta) \right\rangle_{\!(\beta_i, K)}.
\label{eq:zmprew}
\end{eqnarray}
The right-hand side is just the naive summation of $G(\hat{P};\beta,\vec\beta)$ observed in all of the configurations disregarding the differences in the simulation parameters.
The partition function at $(\beta_i, K)$ can be determined, up to an overall factor, by the consistency relations 
\begin{eqnarray}
Z_i 
=\sum_{k=1}^{N_{\rm sim}} N_k \left\langle G(\hat{P};\beta_i,\vec\beta) \right\rangle_{\! (\beta_k, K)}
=\sum_{k=1}^{N_{\rm sim}} N_k \left\langle 
\frac{e^{6N_{\rm site} \beta_i \hat{P}}}{
\sum_{j=1}^{N_{\rm sim}} N_j \, Z^{-1}_j e^{6N_{\rm site} \beta_j \hat{P}} } \right\rangle_{\! (\beta_k, K)}
\end{eqnarray}
for $i=1,\cdots,N_{\rm sim}$. 
Denoting $f_i=-\ln Z_i$, these equations can be rewritten as
\begin{eqnarray}
1 = \sum_{k=1}^{N_{\rm sim}} N_k \left\langle
\frac{1}{ \sum_{j=1}^{N_{\rm sim}} N_j \exp[ 6N_{\rm site} (\beta_j-\beta_i) \hat{P} 
- f_i +f_j]} \right\rangle_{\! (\beta_k, K)},
\hspace{5mm}
i=1,\cdots,N_{\rm sim}.
\label{eq:consis}
\end{eqnarray}
Starting from appropriate initial values of $f_i$, we solve these equations numerically using an iterative method. 
Note that, in these calculations, one of the $f_i$'s must be fixed to remove the ambiguity corresponding to the undetermined overall factor.

The histogram of $|\Omega|$ is obtained using the integral of $w(P,|\Omega|;\beta, K)$ in terms of $P$,
\begin{eqnarray}
w(|\Omega|;\beta, K)
=\sum_{i=1}^{N_{\rm sim}} N_i \left\langle \delta (|\Omega| - |\hat{\Omega}|) 
G(\hat{P};\beta,\vec\beta) \right\rangle_{\!(\beta_i, K)}.
\end{eqnarray}
The expectation value of an operator $\hat{\cal O}[\hat{P},|\hat{\Omega}|]$ at $(\beta, K)$ can be evaluated as
\begin{eqnarray}
\langle \hat{\cal O} [\hat{P},|\hat{\Omega}|] \rangle_{(\beta, K)} 
= \frac{1}{Z(\beta,K)} \sum_{i=1}^{N_{\rm sim}} N_i \left\langle \hat{\cal O} [\hat{P},|\hat{\Omega}|] \, G(\hat{P};\beta,\vec\beta) \right\rangle_{\!(\beta_i, K)},
\label{eq:multibeta}
\end{eqnarray}
where $\hat{\cal O}[\hat{P},|\hat{\Omega}|]$ is an operator that is written using $(\hat{P}, |\Omega|)$, and $ Z(\beta,K)$ is given by Eq.~(\ref{eq:zmprew}).

When one also wants to change the hopping parameters from $K_0$ to $K$, the probability distribution function can be evaluated as 
\begin{eqnarray}
Z^{-1}(\beta, K) \, w(|\Omega|; \beta, K)
= \frac{\sum_{i=1}^{N_{\rm sim}} N_i \left\langle \delta(|\Omega| - |\hat{\Omega}|) 
\prod_f \frac{ \det M(K_f)}{\det M(K_{0;f})} \, 
G(\hat{P};\beta,\vec\beta) \right\rangle_{\! (\beta_i, K_0)} 
}{\sum_{i=1}^{N_{\rm sim}} N_i \left\langle 
\prod_f \frac{ \det M(K_f)}{\det M(K_{0;f})} \, 
G(\hat{P};\beta,\vec\beta) \right\rangle_{\! (\beta_i, K_0)} } ,
\end{eqnarray}
where $G$'s on the right-hand side are defined at $K_0$.
Again, $\sum_{i=1}^{N_{\rm sim}} N_i \left\langle \cdots G \right\rangle_{(\beta_i,  K_0)}$ 
on the right-hand side is just the naive sum of $\cdots G$ over all the configurations disregarding the differences in the simulation parameters. 
A similar expression holds for $\langle \hat{\cal O} [\hat{P},|\hat{\Omega}|] \rangle_{(\beta, K)}$.

\begin{figure}[tb]
\begin{minipage}{0.45\hsize}
\begin{center}
\includegraphics[width=7.5cm]{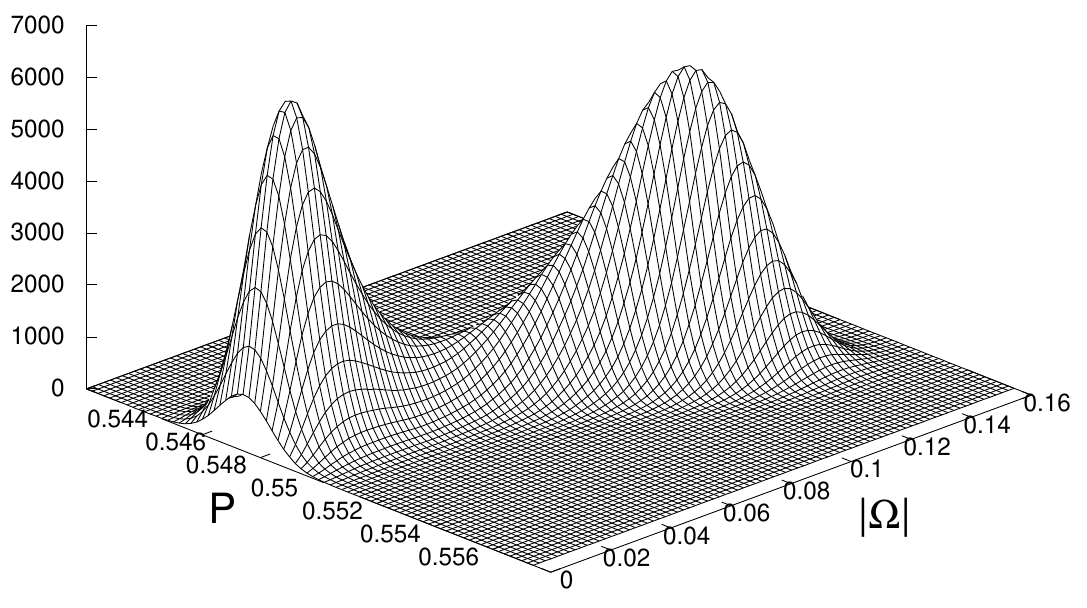}
\end{center}
\caption{The histogram of $P$ and $|\Omega|$ in the heavy quark limit at the transition point, 
calculated using the multi-point histogram method on a $24^3\times4$ lattice.}
\label{fig:24x4histo}
\end{minipage}
\hspace{3mm}
\begin{minipage}{0.45\hsize}
\begin{center}
\includegraphics[width=7.5cm]{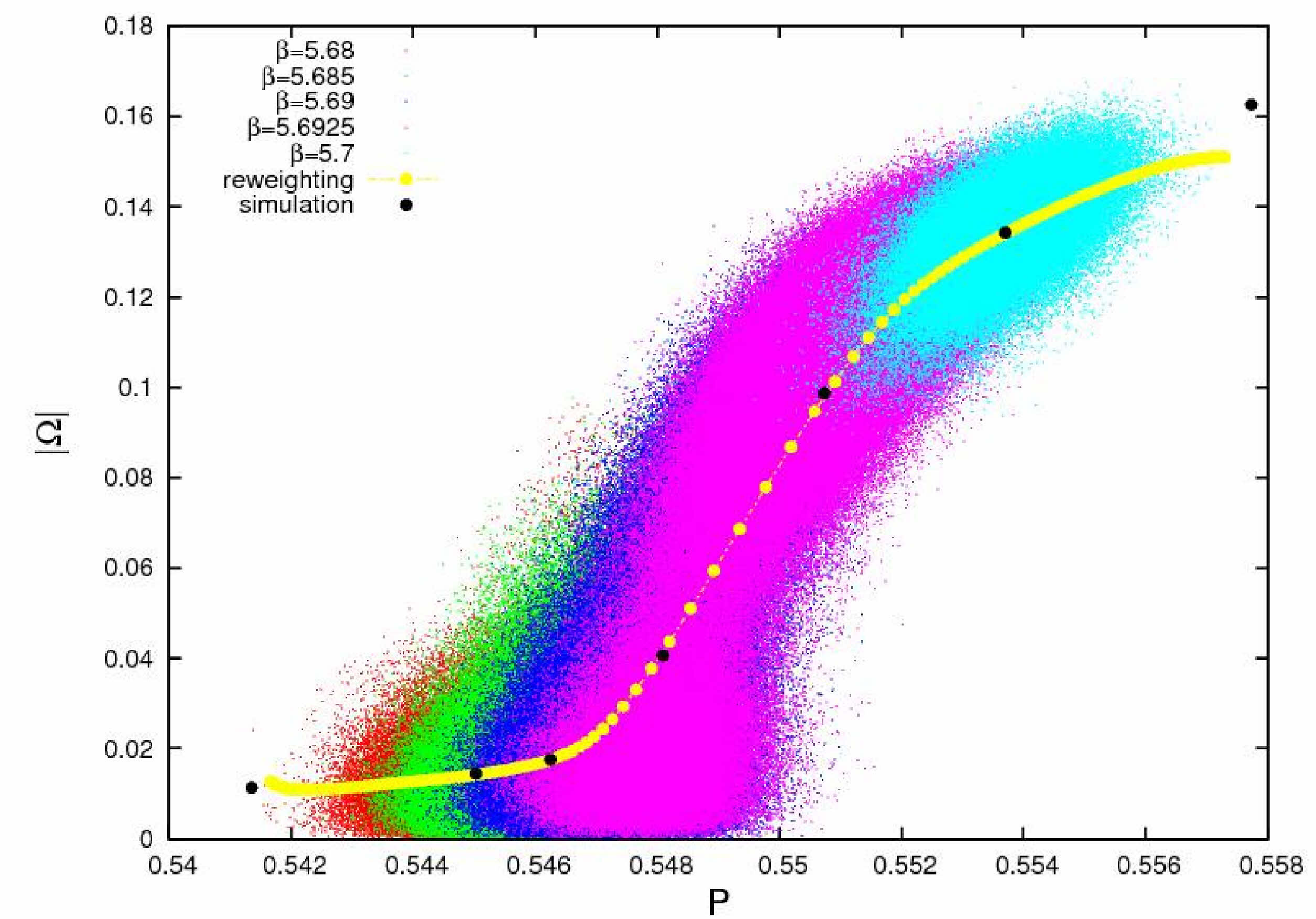}
\end{center}
\caption{Original distribution in the $P$ and $|\Omega|$ plane generated at 5 $\beta$ points on a $24^3\times4$ lattice.
The yellow circles denote the expectation values of $P$ and $|\Omega|$ for $5.66 < \beta < 5.76$.
The black circles are the results from the direct calculation.}
\label{fig:24x4plaqpoly}
\end{minipage}
\end{figure}

Figure \ref{fig:24x4histo} shows an example of the histogram $w(P, |\Omega|;\beta)$
in the heavy quark limit $K=0$ just at the transition point of the pure gauge theory, $\beta = 5.69121$, obtained on a $24^3\times4$ lattice.
Using configurations of Ref.~\cite{Saito1}, data at five different simulation points are combined using the multi-point histogram method. 
We see a double-peak distribution at $K=0$.
The values of $P$ and $|\Omega|$ of each configuration are plotted in Fig.~\ref{fig:24x4plaqpoly}: red, green, blue, purple, and cyan dots represent results at $\beta=5.68$, 5.685, 5.69, 5.6925 and 5.70, respectively.
The yellow circles are the expectation values calculated using the multi-point histogram method varying $\beta$ from 5.66 to 5.76, using the data at $\beta=5.68$--5.70.  
To test the reliable range of the multi-point histogram method, we perform additional simulations at $\beta=5.665$ and $5.76$.
The black circles are the expectation values of $P$ and $|\Omega|$ directly calculated without the reweighting method at $\beta=5.665$ and $5.76$, in addition to the above-mentioned five $\beta$'s.
We find that the results of the direct simulations at $\beta=5.665$ and $5.76$ deviate from the results of reweighting using the data at $\beta=5.68$--5.70. 
These data points are in the regions where the distribution of $P$ and $| \Omega |$ used in the reweighting does not overlap with the target expectation values.
Because the reweighting method only changes the weight of each configuration in the average over the configurations, the method fails when the important region of relevant operators is outside of the original distribution.

\end{document}